\documentclass[a4paper, 12pt]{article}

\usepackage[sort&compress]{natbib}
\bibpunct{(}{)}{;}{a}{}{,} 

\usepackage{amsthm, amsmath, amssymb, mathrsfs, multirow, url, booktabs, float}
\usepackage{graphicx}
\usepackage{caption, subcaption}
\usepackage{ifthen} 
\usepackage{amsfonts}
\usepackage[usenames]{color}
\usepackage{fullpage}
\usepackage[utf8]{inputenc}
\usepackage{xcolor}
\usepackage[normalem]{ulem}
\usepackage[ruled,vlined]{algorithm2e}
\usepackage{soul}
\usepackage{color}

\newcommand{\hlgray}[1]{{\sethlcolor{lightgray}\hl{#1}}}

\theoremstyle{plain}

\newtheorem{proposition}{Proposition}

\theoremstyle{definition}

\theoremstyle{remark} 

\newtheorem{remark}{Remark}

\newtheorem*{assumption*}{Assumption}

\newcommand{\nm}{{\sf N}}

\newcommand{\eps}{\varepsilon}

\title{Valid predictions of random quantities in linear mixed models}

\author{Nicholas Syring\textsuperscript{$\ddagger$,}\footnote{Corresponding author: nsyring@iastate.edu},\quad Fernando Miguez\footnote{Department of Agronomy, Iowa State University}, \quad Jarad Niemi\footnote{Department of Statistics, Iowa State University} 
}

\date{\today}
\begin{document}  
\maketitle

\begin{abstract}   
In applications of linear mixed-effects models, experimenters often desire uncertainty quantification for random quantities, like predicted treatment effects for unobserved individuals or groups.  For example, consider an agricultural experiment measuring a response on animals receiving different treatments and residing on different farms.  A farmer deciding whether to adopt the treatment is most interested in farm-level uncertainty quantification, for example, the range of plausible treatment effects predicted at a new farm.  The two-stage linear mixed-effects model is often used to model this type of data.   However, standard techniques for linear mixed model-based prediction do not produce calibrated uncertainty quantification.  In general, the  prediction intervals used in practice are not valid---they do not meet or exceed their nominal coverage level over repeated sampling.  We propose new methods for constructing prediction intervals within the two-stage model framework based on an inferential model (IM).  The IM method generates prediction intervals that are guaranteed valid for any sample size.  Simulation experiments suggest variations of the IM method that are both valid and efficient, a major improvement over existing methods. We illustrate the use of the IM method using two agricultural data sets, including an on-farm study where the IM-based prediction intervals suggest a higher level of uncertainty in farm-specific effects compared to the standard Student-$t$ based intervals, which are not valid. 
\smallskip

\end{abstract}

\section{Introduction}
\label{S:intro}
Linear mixed effects models are appropriate for a wide range of experiments involving random sampling of and within groups of experimental units.  Common agricultural applications include on-farm crop yield trials across farms and livestock trials across pens or barns--- two examples we analyze below---but the same methodology is used in ecology, medicine, and the social sciences.  Traditionally, inferences based on these models have mainly concerned an overall or population-level treatment effect.  However, from the point of view of a group- or individual-level actor the group- or individual-level mean treatment effect is most relevant.  For example, a patient who did not participate in the trial is more interested in the predicted treatment effect for a new individual with specific covariate values, rather than on the population-level treatment effect because the former is more relevant to individual-level decision-making.  As discussed in \citet{altman.2013} and \citet{altman.2018}, practitioners may struggle to recognize the differences in variability between population-, group-, and individual-level parameters, and do not always choose the appropriate inference method for the parameter of interest.  As pointed out in \citet{higgins.etal.2009} and \citet{inthout.etal.2016}, confidence intervals for overall treatment effect are often used to make inferences on group-level effects, but these intervals systematically underestimate variability at the group level.  Prediction intervals for group-level effects---and not confidence intervals for the overall effect---are appropriate for group-level inferences.

Several methods are available for computing prediction intervals in mixed models, including intervals based on a Student's $t$ approximation to the sampling distribution of the studentized group-level treatment effect, bootstrap-based prediction intervals, and Bayesian prediction intervals.  In a simulation study we find all of these standard prediction intervals experience under-coverage in some cases.  The under-coverage phenomenon for certain Student's $t$ prediction intervals is well-documented in the literature.  For instance, \citet{higgins.etal.2009} suggests a Student's $t$ interval with degrees of freedom equal to the number of groups minus two.  This heuristic was proposed for use in meta-analyses where the lack of raw data makes it challenging to choose the degrees of freedom that yields the best approximation of the sampling distribution.  Several authors \citep{inthout.etal.2016, partlett.2016, laurent.2020} observe that in applications exhibiting very low between-group variability these prediction intervals are not valid. \citet{Francq.etal.2019} propose the same prediction interval for general linear mixed models with degrees of freedom determined by a generalized Satterthwaite approximation.  Alternatively, bootstrap-based predictions may be the most common due to their accessibility in statistical software \citep{lme4}.  Bootstrapping mixed models may be computationally expensive, and \citet{knowles.2020} address this problem with their merTools R package for fast approximation of bootstrap prediction intervals.  Prediction is straightforward from a Bayesian point of view, and, like the bootstrap, Bayesian prediction intervals for mixed models are easily accessible to practitioners using {\tt R} packages {\tt rstanarm} \citep{Goodrich.etal.2022} and {\tt brms} \citep{brms}.  Bayesian prediction intervals are not necessarily meant to meet a nominal coverage level over repeated sampling, but practitioners may still assign them such an interpretation.  Similarly to bootstrap, we found Bayesian prediction intervals, with either the default or a customized choice of prior distributions, did not reliably cover in simulations of random-intercept models.  

Since standard prediction intervals perform poorly in practically relevant examples, the question is: what alternative method produces {\em valid} prediction intervals---ones reliably attaining their nominal coverage level?  In answer to this question we propose prediction intervals based on an {\em inferential model} (IM) following the works of \citet{martin.liu.book, martin.lingham.2016, cella.martin.2020}.  The IM method is model-based and relies heavily on sufficient statistics, so different types of mixed models require different IM methods. We choose to focus on the two-stage mixed model, which is applicable in to the experiments we have in mind. Nevertheless, the same ideas presented herein could be used with other types of mixed models.  A general theory of IM prediction is presented in \citet{cella.martin.2020} where the authors provide sufficient conditions for validity of IM-based prediction intervals in parametric problems.  The two-stage model fits into their setup nicely (see Section 4.3 below) which implies IM prediction intervals based on the two-stage model are valid for any sample size (not just asymptotically).  Provable validity comes at the cost of some efficiency, and simulation results reflect the standard IM approach to be conservative.  Therefore, we suggest two strategies to modify the IM approach to gain efficiency.  In a simulation study, we find the IM approach is the only consistently valid method, and that our suggested modifications increase efficiency without sacrificing practical validity.    

The paper is laid out as follows.  Section~\ref{s:model} introduces the well-known two-stage model.  Section~3 provides a gentle introduction to IM construction and prediction for independent and identically distributed (iid) normal responses.  Section~\ref{S:IM} constructs IM prediction intervals for the two-stage model.  Section~\ref{S:simulations} provides an overview of our extensive simulation study comparing IM prediction intervals to several competing methods in the context of a random-intercept model. Section~\ref{S:examples} includes two real-data agricultural examples.  Section~\ref{s:conclusion} provides concluding remarks.  The appendices include technical details related to IM construction as well as additional simulation results.  Codes for implementing our approach are available in a downloadable {\tt R} package at \verb|https://github.com/nasyring/impred|.

\section{Two-stage linear mixed model}
\label{s:model}
Consider the following Gaussian linear mixed model with two variance components \citep{davidian} often referred to as the two-stage model:
\[
    Y_i = X_i\beta + Z_i\alpha_i + \eps_i, \quad i=1, \ldots, N
\]
where $Y_i$ is an $n_i\times 1$ response vector, $X_i$ an $n_i\times p$ design matrix of covariates, $Z_i$ is an $n_i\times a$ design matrix of covariates, $\beta$ is the $p\times 1$ fixed effects coefficient vector, $\alpha_i$ is an $a\times 1$ normal random vector of random effects with mean zero and covariance matrix $\sigma_\alpha^2 A$ where $A$ is a known $a\times a$ matrix, and $\eps_i$ is an $n_i\times 1$ normal random vector with mean zero and covariance matrix $\sigma^2_\eps I_{n_i}$ such that $(\alpha_i, \eps_i)$ are independent.  For $i = 1, \ldots, N$, $\alpha_i$ and $\eps_i$ are independent sequences of random vectors so that responses are independent between groups.  This model can be used to describe experiments with a hierarchical sampling structure in which groups $i=1, \ldots, N$ are sampled from a population of groups and, subsequently, individuals with responses $Y_{ij}$, $j=1, \ldots, n_i$, are sampled independently from within each group for a total sample size of $n=\sum_{i=1}^N n_i$.  

Let $G$ be an $n\times n$ diagonal block matrix composed of $n_i\times n_i$ blocks $G_i:=Z_iAZ_i^\top$.  An alternative matrix-vector formulation of the two-stage model is as follows
\begin{equation}
    \label{eq:model}
    Y = X\beta + (\sigma_\eps^2 I_n + \sigma_\alpha^2G)^{1/2}U, \quad U\sim \mathsf{N}_n(0, I_n),
\end{equation}
where $M^{1/2}$ denotes the lower Cholesky factor of a matrix $M$. 

The quantity of interest for prediction is the linear combination $\theta = x^\top \beta + z^\top \alpha^\star$ where $x$ and $z$ are covariate vectors corresponding to an unobserved response and where $\alpha^\star~\sim \mathsf{N}_a(0,\sigma_\alpha^2 A)$ is the random effect corresponding to a new group not sampled in the experiment, and hence independent from $\alpha$.  Typically, $\theta$ represents a group-averaged treatment effect given a fixed covariate.  Additionally, we may be interested in predicting a new response from a new group, written $Y^\star = \theta+ \eps^\star$ where $\eps^\star\sim \mathsf{N}(0, \sigma_\eps^2)$ is independent from $\theta$ and $\eps_i$, for all $i=1, \ldots, n$.  Throughout, we use $\theta$ to denote the above random variable and $\vartheta$ to denote a value of this random variable.     

This two variance component model described above is widely applicable and appropriate for the examples we discuss in Sections~\ref{SS:soydata} and \ref{SS:example2}.  However, \eqref{eq:model} only covers linear mixed models with (groupwise) compound symmetric covariance structures.  In principle, there is no reason the IM framework discussed in Sections~\ref{S:illustration} and \ref{S:IM} could not be applied to models with more flexible covariance structures.  As we will explain, a necessary ingredient for (efficient) IM construction is the minimal sufficient statistic; and, so long as it is available we can construct model-based IM predictions using the same ideas as presented below in Section~\ref{S:IM}. 
\section{An illustration of IM prediction}
\label{S:illustration}
In this section, we illustrate the standard three-step method for constructing an IM to predict a normal random variable from iid data.  Our intention is to elucidate how the IM approach to prediction works in a simple example before we tackle the more challenging problem of prediction using the two-stage model. We construct an IM that produces valid prediction intervals for predicting a future response $Y^\star\sim \nm(\mu, \nu^2)$ based on a random sample of size $n$ for unknown $(\mu, \sigma^2)$.  We say a $100(1-\alpha)\%$ prediction interval is valid if it has frequentist coverage probability greater than or equal to the nominal level of $1-\alpha$ for any sample size.

IM construction proceeds in three basic steps: 1) {\em associate} the data, prediction, and an auxiliary random variable with a known distribution via a data-generating equation; 2) {\em predict} the auxiliary random variable with a valid plausibility contour $\pi$; and 3) {\em combine} the contour and association to determine a data-dependent plausibility contour $\pi_n$ for the target.  For further details, see \cite{martin.lingham.2016} and \cite{cella.martin.2020} . 

The first---and often most challenging---step in IM construction is to define an appropriate association, or data-generating equation like that in \eqref{eq:model}.  We start with $n+1$ data-generating equations for the observations $Y^n = (Y_1,\ldots,Y_n)^\top$ and future scalar observation $Y^\star$: 
\[Y^n = \mu + \nu I_n \Psi^n,\quad   Y^\star = \mu + \nu \Psi^\star,\]
where $\Psi^n = (\Psi_1, \ldots, \Psi_n)^\top$, $\Psi_j \stackrel{iid}{\sim}N(0,1)$ for $j=1, \ldots, n$, and $\Psi^\star\sim N(0, 1)$, independent from $\Psi^n$.  The idea is to use the association like a system of equations that we can solve to determine the values of the unknown parameters.  Ideally, we would substitute the observations $Y^n = y^n$ and samples $\Psi^n = \psi^n$ and $\Psi^\star = \psi^\star$ of the auxiliary random variables into the association equations and solve for $(\mu, \nu, y^\star)$.  However, in order to yield a unique solution the number of equations in the association should match the dimension of the parameter vector, and, unfortunately, we have $(n+1)$ equations and only $3$ unknowns.  \cite{martin.liu.book} discusses reducing the dimension of associations, and suggests rewriting the association so that it depends on the data only through the minimal sufficient statistic and its sampling distribution.  Further dimension-reduction techniques focus on removing unnecessary associations involving only nuisance parameters, here $\mu$ and $\nu$.  Using the sample mean $\overline Y_n$ and the sample variance $S_n^2$ we have the three-dimensional association
\begin{align}
\label{eq:assoc.ex1}
S_n^2 = \frac{\nu^2}{n-1}\chi^2,\,\,\,\,\,
\overline Y_n = \mu + \tfrac{\nu}{\sqrt{n}}I_n\Psi^n,\,\,\,\,\,\text{and} \,\,
Y^\star = \mu +  \nu \Psi^\star.
\end{align}
Solve for $(\mu,\nu)$ in the first two equations above and substitute into the third to obtain
\begin{equation}
    \label{eq:assoc.ex}
    Y^\star =  \overline Y_n +  T\sqrt{S_n^2(1+\tfrac{1}{n})},
\end{equation}
where $T\sim T_{n-1}$ has a Student t distribution with $n-1$ degrees of freedom. Such substitutions are justified by the IM principle of {\em marginalization}, by which we may drop unnecessary associations after substitution.  In this case, we keep only the association in (3) while dropping the associations for the nuisance parameters in (2).  The reasoning is as follows: for any $(Y^\star, \overline Y_n, T, S^2_n)$ satisfying \eqref{eq:assoc.ex} there is a pair $(\nu^2, \mu)$ that solves the first two equations in \eqref{eq:assoc.ex1}. These are free variables that do not carry any information about $Y^\star$, so we may safely ignore/marginalize those two equations.       

The next step is to choose a plausibility contour $\pi(t)$ for predicting the auxiliary random variable $T$.  The function $\pi(t)$ may be any function mapping the domain of $T$ to $[0,1]$. But, in order to obtain valid prediction intervals the auxiliary contour must satisfy the following {\em validity} property: for all $\alpha\in (0,1)$,
\begin{equation}
    \label{eq:plaus.optim}
    P_T\{\pi(T)\leq \alpha\}\leq \alpha, \quad T\sim T_{n-1}.
\end{equation}
In other words, $\pi(T)$ is stochastically no smaller than a uniform random variable, with respect to $T\sim T_{n-1}$.  At least in this example it turns out that an optimal choice of $\pi(t)$ is available --- one that leads to the most efficient inferences about $Y^\star$, e.g., tightest valid prediction intervals --- and it is given by 
\[\pi(t) = P_T\{f(T)\leq f(t)\}, \quad T\sim T_{n-1}\]
where $f$ is the Student t density function for $n-1$ degrees of freedom; and see \cite{martin.liu.2020} for more on this so-called {\em maximum-specificity contour}.   

For the final step we combine the plausibility contour for $T$ with the association in \eqref{eq:assoc.ex} to derive a plausibility contour for $Y^\star$.  Given a predicted value $y^\star$ and the observed data $y^n$ write $t_y:=(y^\star - \overline y_n)/\sqrt{s_n^2(1+1/n)}$ for the solution in $T$ to \eqref{eq:assoc.ex} where $\overline y_n$ and $s_n^2$ are the observed sufficient statistics; then, the plausibility contour for $Y^\star$ is defined by $\pi_n(y^\star) = \pi(t_y)$.  

Define a $100(1-\alpha)\%$ prediction interval for $Y^\star$ by the $\alpha-$cut $C_\alpha(y^n):=\{y^\star:\pi_n(y^\star)>\alpha\}$ of $\pi_n(y)$, which equals the following
\begin{align*}
    C_\alpha(y^n)=&\left\{y:P_T\left(f(T)<f\left(t_y\right)\right)>\alpha\right\}.
\end{align*}
Let $T_{m,\alpha}$ denote the $\alpha^{th}$ quantile of Student's $t$ distribution with $m$ degrees of freedom.  Then, $\{z:P_T(f(T)<f(z))>\alpha\}$ is simply $\{z: T_{n-1, \alpha/2}\leq z\leq T_{n-1, 1-\alpha/2}\}$, and it follows that $C_\alpha(y^n)$ is equivalent to the interval
\begin{align*}\overline y_n \pm T_{n-1, 1-\alpha/2}\sqrt{s_n^2(1+1/n)},
\end{align*}
which is the classical (exactly) valid prediction interval for $Y^\star$ \citep{fisher.1935} and the Bayesian prediction interval based on the default prior $1/\sigma$.  But, we need not rely on existing results to show validity of IM-based prediction intervals. Rather, general IM theory for prediction is available to prove such results; and, see Section~\ref{SS:validity}.

The IM framework may be unfamiliar to most readers, but as the above example shows, its predictions coincide with those of standard procedures in simple problems.  As we show in Section~\ref{S:IM}, the advantage of the IM framework is its ability to produce valid prediction intervals in more challenging settings where standard methods fall short.

\section{An IM for the two stage model}
\label{S:IM}
In this section we develop two different IMs for predicting $\theta$.  In Sections~4.1-2 we apply the same three-step construction used in Section~\ref{S:illustration}, but in the case of the two-stage model, IM marginalization cannot completely remove nuisance parameters.  Instead, following the standard IM construction leads to a joint IM for the two-dimensional parameter $(\theta, \rho)$ where $\rho = \sigma_\alpha^2(\sigma_\alpha^2 + \sigma_\eps^2)^{-1}$ is often referred to as the intra-class correlation (or heritability) coefficient.  Marginal prediction of $\theta$ based on the joint IM is not efficient, so in Section 4.3 we propose a generalized marginal IM strategy based on the ideas in \citet[Chapter 7.4]{martin.liu.book}.  Section 4.4 includes a proof of the validity of joint IM prediction intervals, and discusses modifications of both IM strategies to reduce the average width of prediction intervals.

\subsection{Association step}
\label{ss:associate}
Begin with the data-generating equation in \eqref{eq:model}:
\begin{align*}
        Y &= X\beta + (\sigma_\eps^2 I_n + \sigma_\alpha^2G)^{1/2}U, \quad U~\sim \mathsf{N}_n(0, I_n),
\end{align*}
which includes $n$ equations, one for each response.  Our first goal is to use minimal sufficient statistics to reduce the number of association equations as much as possible, ideally to only $p+2$ equations, matching the number of unknown parameters.  

\citet{olsen.1976} show the minimal sufficient statistics for the two-component model are given by $(BY, S_1, \ldots, S_L)$ where $BY =  (X^\top X)^{-1}X^\top Y$ estimates the regression coefficient and $(S_1, \ldots, S_L)$ are sums of squares jointly sufficient for $(\sigma_\alpha^2, \, \sigma_\eps^2)$---their precise definitions are given in Appendix~\ref{A:assoc.suff} \citet{martin.liu.book} define an association for $(\beta, \sigma_\alpha^2, \sigma_\eps^2)$ which we give below with an additional association equation for predicting $\theta$: \begin{equation}
\label{eq:assoc.unreduced}
\begin{aligned}
    S_\ell &= (\lambda_\ell \sigma_\alpha^2 + \sigma_\eps^2)V_\ell,\,\,\,\,\quad\quad\quad\quad V_\ell\stackrel{ind.}{\sim} \chi^2(r_\ell), \quad \ell = 1,\ldots, L;\\
	BY &= \beta + C_\sigma^{1/2} W_1, \,\,\,\,\,\,\,\,\,\quad\quad\quad\quad W_1 \sim \mathsf{N}_p(0, I_p),\\
	\theta &= x^\top\beta + (\sigma_\alpha^2 z^\top A z)^{1/2}W_2, \quad W_2 \sim \mathsf{N}(0, 1)
\end{aligned}
\end{equation}
where $L\geq 2$ is the number of distinct eigenvalues of a known matrix $H$ (see Appendix~\ref{A:assoc.suff}); $\lambda_\ell$ and $r_\ell$ for $\ell = 1, \ldots, L$ are the eigenvalues in decreasing order and their multiplicities, respectively; $C_\sigma = (\sigma_\eps^2 BB^\top + \sigma_\alpha^2 B G B^\top)$ is a $p\times p$ matrix; and, $(W_1, W_2, V_1, \ldots, V_L)$ are independent.  

Following \citet[Chapter 7]{martin.liu.book}, the association in \eqref{eq:assoc.unreduced} is {\em regular} with respect to the nuisance parameter $\beta$.  As a result, we may substitute $\beta$ by $BY - C_\sigma^{1/2}W_1$ in the third line and then ignore/marginalize the second line.  This leaves us with the following $(L+1)-$dimensional association:
\begin{equation}
\label{eq:assoc.unreduced3}
\begin{aligned}
    S_\ell &= (\lambda_\ell \sigma_\alpha^2 + \sigma_\eps^2)V_\ell,\quad V_\ell\stackrel{ind.}{\sim} \chi^2(r_\ell), \quad \ell = 1,\ldots, L;\\
	\theta &= x^\top BY + (x^\top C_\sigma x + \sigma_\alpha^2 z^\top A z)^{1/2} W, \quad W \sim \mathsf{N}(0, 1).
\end{aligned}
\end{equation}
Just as in the illustration in Section~\ref{S:illustration}, marginalization is justified for the following reason: for any combination of auxiliary random variable, parameter, and data values solving the equations in \eqref{eq:assoc.unreduced3}, there always exists a vector $\beta$ simultaneously satisfying the $BY$ equation in \eqref{eq:assoc.unreduced}.  Therefore, the $BY$ equation carries no information about $\theta$ or the variance components and may be ignored after the substitution.

The association in \eqref{eq:assoc.unreduced3} is not regular with respect to the variance components, but, after an appropriate reparametrization and substitution, we may perform one more marginalization step.  Define $\rho = \sigma_\alpha^2(\sigma_\alpha^2+\sigma_\eps^2)^{-1}$ and rewrite \eqref{eq:assoc.unreduced3} using $(\sigma_\alpha^2, \sigma_\eps^2)\mapsto(\rho, \sigma_\eps^2)$ by dividing by $S_L$ in the association equations for $S_\ell$, $\ell \ne L$, and $\theta$.  The result is a regular association with respect to the nuisance parameter $\sigma_\eps^2$ involving the equation $S_L = \sigma_\eps^2[\lambda_L\rho(1-\rho)^{-1}+1]V_L$, which is marginalized/dropped.  We are left with the $L-$dimensional association for $(\theta, \rho)$:
\begin{equation}
\label{eq:assoc.reduced}
\begin{aligned}
&\frac{S_\ell}{S_L} = \frac{\rho(\lambda_\ell-1)+1}{\rho(\lambda_L-1)+1}\frac{V_\ell}{V_L}, \quad \ell = 1, \ldots, L-1;\\
&\frac{\theta - x^\top BY}{S_L^{1/2}}\left(\frac{\rho(\lambda_L-1)+1}{\rho(c_1 - 1)+c_2}\right)^{1/2} = \frac{W}{V_L^{1/2}},
\end{aligned}
\end{equation}
where $c_1 = x^\top BB^\top x$ and $c_2 = z^\top (Z+BGB^\top)z$.

We have pushed the IM marginalization strategies as far as we can; \eqref{eq:assoc.reduced} is not regular, so we seem to be stuck with $L$ equations for two parameters, including the nuisance parameter $\rho$.  In some cases---like random intercept models for balanced experiments---$L=2$ so that \eqref{eq:assoc.reduced} has exactly two equations for two parameters; with no further marginalization possible we cannot do any better.  However, for most applications $L>2$ so that \eqref{eq:assoc.reduced} contains more than one association involving only the parameter $\rho$.  For those cases, \citet{martin.hc} implemented a so-called local-conditional association for $\rho$ that reduces $L-1$ association equations involving only $\rho$ down to just one; see \citet[Chapter 8.3.4]{martin.liu.book} and/or \citet{martin.hc} for details.  Their association depends on a given value $\rho = \rho_0$.  Like a null distribution, their local association is correctly-specified only when $\rho=\rho_0$; so, it can be used to evaluate point-null hypotheses about $\rho$, and, importantly, to define valid $100(1-\alpha)\%$ confidence intervals for $\rho$ by collecting all such point-null values with plausibility (p-value) above $\alpha$.  Next, we augment their local conditional association for $\rho$ with the equation for $\theta$ in \eqref{eq:assoc.reduced} to derive the following two-dimensional association for $(\theta, \rho)$, which we need for applications with $L>2$:
\begin{equation}
\label{eq:assoc.local.cond}
\begin{aligned}
&\sum_{\ell = 1}^{L-1} \log\left[\frac{S_\ell}{S_L}\right] = \sum_{\ell = 1}^{L-1} \log\left[\frac{\rho_0(\lambda_\ell-1)+1}{\rho_0(\lambda_L-1)+1}\right] + U_0;\\
&\frac{\theta - x^\top BY}{S_L^{1/2}}\left(\frac{\rho_0(\lambda_L-1)+1}{\rho_0(c_1 - c_2)+c_2}\right)^{1/2} = \frac{W}{V_L^{1/2}},
\end{aligned}
\end{equation}
where $U_0$ has the distribution of $\sum_{\ell = 1}^{L-1} \log\left[\frac{V_\ell}{V_L}\right]$ conditioned on the linear combination \begin{align*}
    &\left(\log \left[V_1 / V_L\right], \ldots, \log \left[V_1 / V_{L-1}\right]\right)^\top M_0 \\
    & \,\,= \left(\log\left[\frac{S_1}{S_L}\right] - \log\left[\frac{\rho_0(\lambda_1-1)+1}{\rho_0(\lambda_L-1)+1}\right], \ldots, \log\left[\frac{S_{L-1}}{S_L}\right] - \log\left[\frac{\rho_0(\lambda_{L-1}-1)+1}{\rho_0(\lambda_L-1)+1}\right]\right)^\top M_0.
\end{align*} 
for a known, fixed, $(L-1)\times(L-2)-$dimensional matrix $M_0$ depending only on $\rho_0$; and see the Appendix~\ref{A:assoc.suff} for technical details.     

We have now completed the association step of IM construction.  \eqref{eq:assoc.reduced} and \eqref{eq:assoc.local.cond} provide two-dimensional associations for prediction/inference of $(\theta, \rho)$ for the two cases $L=2$ and $L>2$.  In the next section we complete IM construction for the more common case of $L>2$ using the association in \eqref{eq:assoc.local.cond} by applying the IM predict and combine steps (the $L=2$ case is simpler, and can be handled similarly).  Two strategies---a joint IM and a generalized IM strategy---are available for completing IM construction and producing valid prediction intervals for $\theta$.  The joint IM strategy uses the full association in \eqref{eq:assoc.local.cond} to construct simultaneous, valid prediction/confidence regions for $(\theta, \rho)$.  These may be projected to the domain of $\theta$ to produce valid (albeit conservative) prediction intervals for $\theta$.  The generalized IM strategy aims to deliver less conservative prediction intervals for $\theta$ compared to the joint IM.  The idea is to develop a one-dimensional association for $\theta$ that is valid for any values of the variance components $(\sigma_\alpha^2, \sigma_\eps^2)$. Reducing the dimension of the association should reduce overcoverage of prediction intervals, but the requirement the association is valid for all values of variance components---a requirement needed for validity---will tend to make the prediction intervals conservative.  The next two sections detail these two strategies, and the simulation experiments in Section~\ref{S:simulations} provide some guidance as to which is most efficient.
\subsection{Predict and combine steps for the joint IM}
\label{SS:predict_combine}
Given the association in \eqref{eq:assoc.local.cond} for the $L> 2$ case we move on to the {\em predict} and {\em combine} steps where we apply the maximum specificity contour based on the joint density $f_{\rho_0}(u,v)$ of the auxiliary random variables appearing in \eqref{eq:assoc.local.cond}; and see Appendix~\ref{A:assoc.suff}.  To complete the IM specification we combine the association in \eqref{eq:assoc.local.cond} with the maximum specificity contour to get the following plausibility contour:
\begin{equation}
    \label{eq:jointcontour1}
    \pi_n(\vartheta, \rho_0) := \pi(U,V) = P\left(f_{\rho_0}(U', V') < f_{\rho_0}\left(U, V\right)\right),
\end{equation}
where $(U',V')$ are random variables with joint density $f_{\rho_0}$, and where
\begin{equation}
\label{eq:UV}
    \begin{aligned}
    U &= \sum_{\ell = 1}^{L-1} \log\left[\frac{S_\ell}{S_L}\right] - \sum_{\ell = 1}^{L-1} \log\left[\frac{\rho_0(\lambda_\ell-1)+1}{\rho_0(\lambda_L-1)+1}\right], \text{ and}\\
    V &= \frac{\vartheta - x^\top BY}{S_L^{1/2}}\left(\frac{\rho_0(\lambda_L-1)+1}{\rho_0(c_1 - c_2)+c_2}\right)^{1/2}.
    \end{aligned}
\end{equation}

The contour $\pi_n(\vartheta, \rho_0)$ produces valid p-values for the hypotheses $H_0:\{\theta = \vartheta, \, \rho = \rho_0$\}, and the sets $\{(\vartheta, \, \rho_0): \pi_n(\vartheta, \rho_0) > \alpha\}$ constitute valid $100(1-\alpha)\%$ simultaneous prediction/confidence sets; and, see Section~\ref{SS:validity} below.  Projecting these sets to $\theta$ yields valid prediction intervals given by $\{\vartheta: \pi_n(\vartheta, \rho_0) > \alpha\}$.  Equivalently, we may compute the marginal contour
\begin{equation}
    \label{eq:jointcontour}
    \pi_n^J(\vartheta) = \sup_{\rho_0}\pi_n(\vartheta, \rho_0),
\end{equation}
where the superscript $J$ denotes the contour is derived from the joint IM, and define the $100(1-\alpha)\%$ prediction interval for $\theta$ to be the set $\{\vartheta: \pi_n^J(\vartheta)>\alpha\}$.  Computation of $\pi_n^J(\vartheta)$ is straightforward given MCMC samples from $f_{\rho_0}$; and, see Algorithm~\ref{algo1}.
\begin{algorithm}[h]
\SetAlgoLined
 Choose a large integer $M>0$, an equally-spaced grid $\rho_1, \ldots, \rho_J$ in $(0,1)$, and a value $\vartheta$.\\
 \For{$j = 1, \ldots, J$}{1. Compute the realized auxiliary random variables:
 \[u = \sum_{\ell = 1}^{L-1} \log\left[\frac{s_\ell}{s_L}\right] - \sum_{\ell = 1}^{L-1} \log\left[\frac{\rho_j(\lambda_\ell-1)+1}{\rho_j(\lambda_L-1)+1}\right],\]
and,
\[v = \frac{\vartheta - x^\top By}{S_L^{1/2}}\left(\frac{\rho_j(\lambda_L-1)+1}{\rho_j(c_1 - c_2)+c_2}\right)^{1/2}.\]
2. Compute the density of the realized auxiliary random variables $f_{\rho_j}(u,v)$.\\
 \For{$m=1, \ldots, M$}{ 1. Sample $(U'_m,V'_m) \sim F_{\rho_j}$. \\
 2. Store the density values $f_{\rho_j}(U'_m,V'_m)$.  \\
3. Approximate the plausibility of $(\vartheta, \rho_j)$ by
\[\hat\pi_n(\vartheta, \rho_j) = M^{-1}\sum_{m=1}^m{1\left\{f_{\rho_j}(U'_m,V'_m)\leq f_{\rho_j}(u,v)\right\}}.\]
}
}
 \KwResult{$\hat\pi_n^J(\vartheta) = \max_j\pi_n(\vartheta, \rho_j)$.}
 \caption{Monte Carlo approximation of the plausibility contour $\pi_n^J(\vartheta)$.}
 \label{algo1}
\end{algorithm}  

\subsection{Constructing a generalized IM for prediction}

The drawback of the joint IM is that it produces conservative marginal inferences for $\theta$ by point-wise maximization of the joint plausibility function over the intra-class correlation coefficient $\rho$.  For a more efficient approach involving only a one-dimensional association we consider a generalized IM for $\theta$.   

First, we need a one-dimensional association for $\theta$.  Following the developments in Section~\ref{ss:associate} we choose the following association
\begin{equation}
    \label{eq:genIMassoc}
    \frac{(\theta - x^\top BY)\left(\sum_{\ell = 1}^{L-1} r_\ell\right)^{1/2}}{\left( \sum_{\ell = 1}^{L-1} S_\ell \frac{c_1\eta + c_2}{\lambda_\ell \eta + 1}\right)^{1/2}} = t_\nu,
\end{equation}
where $\eta = \sigma_\alpha^2\sigma_\eps^{-2}$ is the variance ratio and $t_\nu$ is a Student's $t$ random variable with $\nu = \sum_{\ell = 1}^{L-1} r_\ell$ degrees of freedom; we will explain why we choose this particular association below.  

Given the true value of $\eta$, \eqref{eq:genIMassoc} is correctly-specified, and we may define a valid plausibility contour for $\theta$ using the maximum specificity auxiliary contour:
\[\pi_n(\vartheta) = \pi(t) = P(f_\nu(T) < f_{\nu}(t)),\]
where $T\sim F_\nu$ is a Student's $t$ random variable with $\nu$ degrees of freedom and
\[t = \frac{(\theta - x^\top By)\left(\sum_{\ell = 1}^{L-1} r_\ell\right)^{1/2}}{\left( \sum_{\ell = 1}^{L-1} s_\ell \frac{c_1\eta + c_2}{\lambda_\ell \eta + 1}\right)^{1/2}} \]
is the observed value of $t_\nu$.
Of course, $\eta$ is unknown, so the contour $\pi_n(\vartheta)$ defined above is of no practical use.  On the other hand, if we let 
\[t' = \frac{(\theta - x^\top By)\left(\sum_{\ell = 1}^{L-1} r_\ell\right)^{1/2}}{\sup_\eta \left( \sum_{\ell = 1}^{L-1} s_\ell \frac{c_1\eta + c_2}{\lambda_\ell \eta + 1}\right)^{1/2}},\]
then $P(f_\nu(T) < f_\nu(t))\leq P(f_\nu(T) < f_\nu(t'))$ for all $\eta$ and the {\em generalized} plausibility contour defined by
\begin{equation}
\label{eq:genIMcontour}
    \pi_n^G(\vartheta) = \pi(t')
\end{equation}
is valid for $\theta$ for any value of $\eta$.

The key property of the association in \eqref{eq:genIMassoc} is that $\sup_\eta \left( \sum_{\ell = 1}^{L-1} S_\ell \frac{c_1\eta + c_2}{\lambda_\ell \eta + 1}\right)^{1/2}$ is finite almost surely, so that the resulting generalized IM plausibility contour does not degenerate to $\pi_n(\vartheta) = 1$ for all $\vartheta$.  In practice, it is often the case $\lambda_L = 0$, in which case the above sum taken over $\ell = 1, \ldots, L$ typically is unbounded when maximized over $\eta$---this is our reason for omitting the $L^{th}$ sufficient statistic $S_L$ from the generalized IM association.

\subsection{Validity of IM-based prediction intervals}
\label{SS:validity}

The fact that the plausibility contour $\pi_n^J(\vartheta)$ in \eqref{eq:jointcontour} developed in Sections~\ref{ss:associate}--\ref{SS:predict_combine} produces valid prediction intervals defined by $C_\alpha(y):=\{\vartheta:\pi_n^J(\vartheta) \geq \alpha\}$ follows from the general theory of IM prediction developed in \citet{cella.martin.2020}.  Among the results they show is that the following condition is sufficient for validity of joint IM prediction intervals:
\begin{equation}
    \label{eq:suff}
    P_{Y, \theta}(\pi_n^J(\theta)\leq \alpha)\leq \alpha \quad \text{for all} \,\, (\alpha, n, \beta, \sigma_\alpha^2, \sigma_\eps^2)
\end{equation}
and see their Proposition 3 and Theorem 1.  To show \eqref{eq:suff} first suppose the true value $\rho$ corresponding to the true values of the variance components is known.  Then, by definition
\begin{align*}
    \pi_n(\theta, \rho) &= \pi(U,V)= P_{U',V'}(f_{\rho}(U',V') < f_{\rho}(U,V))
\end{align*}
where $(U',V') \sim F_\rho$ and $(U, V)\sim F_\rho$ are defined in \eqref{eq:UV}.  Since $(U,V)$ and $(U',V')$ are iid, $\pi(U,V)$ is a uniform random variable with respect to $F_\rho$, or equivalently, with respect to the joint distribution of $(Y,\theta)$, and, as a result, \[P_{Y,\theta}(\pi_n(\theta, \rho) \leq \alpha) = P_{Y,\theta}(\pi(U,V)\leq \alpha ) = \alpha.\]
For the final step, note that, by definition, $\pi_n^J(\theta)\geq \pi_n(\theta, \rho)$ almost surely, so that $P_{Y, \theta}(\pi_n^J(\theta)\leq \alpha) \leq P_{Y,\theta}(\pi_n(\theta, \rho) \leq \alpha) = \alpha$.  Hence, \eqref{eq:suff} is satisfied and as a consequence of Proposition~3 and Theorem~1 in \citet{cella.martin.2020} the following claim concerning the coverage of IM prediction intervals holds.
\begin{proposition}
\label{prop:1}
The $100(1-\alpha)\%$ prediction interval defined by $C_\alpha(y):=\{\vartheta:\pi_n^J(\vartheta)\geq \alpha\}$ satisfies
\[P_{Y, \theta}(\theta \in C_\alpha(Y))\geq 1-\alpha, \quad\text{for all}\,\, (\alpha, n, \beta, \sigma_\alpha^2, \sigma_\eps^2).\]
\end{proposition}
In practice, the plausibility contour $\pi_n^J(\vartheta)$, and, hence, the prediction intervals $C_\alpha(y)$, are approximated by MCMC and maximization over a grid.  Therefore, the validity property is achieved approximately, in a sense, but this approximation is only a function of the number of MCMC samples used and the size and location of the grid, and not the sample size.  So, the approximation error, conceivably, can be made negligible.

Essentially the same argument made above proving validity of the joint IM plausibility contour can be made for the generalized IM plausibility contour given in \eqref{eq:genIMcontour}.  The upshot is that both methods produce valid prediction intervals for $\theta$, but which is ``better" (more efficient)? Intuitively, we expect neither to be efficient, because both must account---in one way or another---for a nuisance parameter, $\rho$ or $\eta$.  However, heuristic modifications to the joint and generalized IM procedures might lead to efficiency gains, but sacrifice guaranteed validity.  For the joint IM contour $\pi_n^J(\vartheta)$ it is reasonable to suspect joint prediction/confidence sets for $(\theta, \rho)$ to behave like set products of a prediction interval for $\theta$ and a confidence interval for $\rho$.  If so, then Bonferroni's argument implies a, say, $90\%$ joint prediction/confidence set corresponds roughly to two crossed $95\%$ intervals.  This suggests using the set $\{\vartheta:\pi_n^J(\vartheta)\geq 0.1\}$ as a $95\%$ prediction interval, which will be shorter than $\{\vartheta:\pi_n^J(\vartheta)\geq 0.05\}$, and might still achieve $95\%$ coverage, due to the over-coverage of the joint IM.  On the other hand, the generalized IM tends to be overly conservative because it is required to be valid for all $\eta$ values, even those that are totally implausible according to the data.  Rather than committing to this ``worst case scenario", we might consider a plausibility contour based on the association in \eqref{eq:genIMassoc} with $\eta$ replaced by a constant, data-dependent value.  Of course, if we set $\eta$ equal to a consistent point estimator $\hat\eta$ the corresponding prediction intervals will be only asymptotically valid.  Setting $\eta$ equal to, say, $\hat\eta \pm \delta$, for some $\delta>0$ and where $\pm$ is determined so as to maximize the denominator in \eqref{eq:genIMassoc}, provides a compromise between the generalized IM contour and the contour based on a plug-in estimate.  A reasonable value of $\delta$ might be the bootstrap standard error of the restricted maximum likelihood estimate of $\eta$, for example.  Both of these heuristic modifications are examined in the simulation experiments in Section~\ref{S:simulations}.

\section{Simulations}
\label{S:simulations}

In this section we investigate the frequentist coverage properties of prediction intervals for several methods in the context of the random intercept model, a special case of \eqref{eq:model}, defined by $Y_{ij} = \mu + \alpha_i + \eps_{ij}$ where $i$ indexes groups and $j$ indexes individuals within group $i$, and where $\alpha_i \sim \mathsf{N}(0, \sigma_\alpha^2)$ is a group-wise random effect. Our target for prediction is $\theta = \mu + \alpha^\star$, representing the average response in a new group; in Appendix~\ref{A:sims} we also include predictions for a new response in a new group, $Y^\star = \theta + \eps$.    

We set $\mu = 0$ and consider twelve scenarios where we vary the values of variance components over the pairs $(\sigma_a^2, \,\sigma^2) = (0.1, 1.0)$, $(0.5, 0.5)$, and $(1.0, 0.1)$ and vary the design over both small and medium, and balanced and unbalanced designs.  Our four designs are:  
\begin{itemize}
    \item[A. ] Balanced, small study with $5$ groups of $6$ observations each.
    \item[B. ] Balanced, medium-sized study with $10$ groups of $12$ observations each.
    \item[C. ] Unbalanced, small study with $3$ groups of $4$ observations, $1$ group with $6$ observations, and $1$ group with $12$ observations.
    \item[D. ] Unbalanced, medium-sized study with $10$ groups of the following sizes: $4$, $4$, $7$, $11$, $13$, $16$, $16$, $16$, $16$, and $17$.
\end{itemize}

We compare our IM prediction intervals to five other methods:
\begin{itemize}
    \item[i) ] \textit{Oracle} method prediction intervals use the true values of the variance components and have endpoints given by \[\overline{y}_n  \pm z_{1-\alpha/2}\left\{\sigma_\alpha^2\left(1+\tfrac{1}{n^2}\sum_{i=1}^i n_i^2\right) + \sigma_\eps^2\tfrac1n \right\}^{1/2}\] where $z_{\alpha}$ is the lower $100\alpha\%$ standard normal quantile and $\overline{y}_n$ is the sample mean response. 
    \item[ii) ] \textit{Student's $t$} prediction intervals have the same form as the Oracle intervals, but with the variance components replaced by their restricted maximum likelihood estimates $(\hat\sigma_\alpha^2, \hat\sigma_\eps^2)$, and the normal distribution quantiles replaced by quantiles of a Student's $t$ distribution with $I-2$ degrees of freedom; as suggested by \cite{higgins.etal.2009}.
    \item[iii) ] \textit{IM} prediction intervals are computed four ways.  Joint $95\%$ intervals are computed using Algorithm~\ref{algo1}. Adjusted $95\%$ intervals are computed as projections of joint $90\%$ prediction confidence sets, as suggested in the comments at the end of Section~\ref{SS:validity}.  In each iteration of Algorithm~\ref{algo1} we use $5000$ MCMC samples and an equally-spaced grid of 100 $\rho$ values between 0.001 and 0.999.  Each simulation run required, on average, $3.5$ seconds.  Generalized IM intervals are computed using the contour $\pi_n^G(\vartheta)$ defined in \eqref{eq:genIMcontour}.  We used 10000 Monte Carlo samples to approximate $\pi_n^G(\vartheta)$.  Adjusted generalized IM prediction intervals are computed using the association in \eqref{eq:genIMassoc} with $\eta$ set equal to its restricted maximum likelihood estimate plus or minus one standard error (computed using 100 bootstrap resamples), as suggested in the comments at the end of Section~\ref{SS:validity}.   Each simulation run required, on average, $2$ seconds with bootstrap, or $0.1$ seconds without bootstrap.   
    \item[iv) ] \textit{Nonparametric bootstrap} prediction intervals for $\theta^\star$ are computed using the percentile method and stratified resampling.  To compute the bootstrap distribution of the within-group means we sample with replacement within each group and return the bootstrapped within-group sample means.  A $100(1-\alpha)\%$ prediction interval for a new group mean is defined by the $\alpha/2$ and $1-\alpha/2$ quantiles of this bootstrap distribution.  Each simulation run required, on average, $3.75$ seconds. 
    \item[v) ] \textit{Parametric bootstrap} prediction intervals for $\theta$ are computed using the {\tt lme4} package and the functions {\tt lmer} and {\tt bootMer}.  These functions implement the parametric bootstrap of the random intercept model.  For each bootstrap-resampled set of responses $y^{n,b} = (y_{11}^b, \ldots, y_{In_i}^b)^\top$ we compute the quantity \[\theta^b = \overline y^{n,b} + z^b\left\{\hat\sigma_\alpha^2\left(1+\tfrac{1}{n^2}\sum_{i=1}^i n_i^2\right) + \hat\sigma_\eps^2\tfrac1n \right\}^{1/2}\] where $z^b\stackrel{iid}{\sim}\mathsf{N}(0,1)$. Repeat for $b=1,\ldots,B$ times and define a $100(1-\alpha)\%$ prediction interval for a new group mean by the $\alpha/2$ and $1-\alpha/2$ quantiles of the values $(\theta^{1}, \ldots, \theta^{B})$.  This method is the most computationally demanding of those we consider, and it is necessary to use only $B=500$ resamples to perform the simulation in a reasonable amount of time; each simulation run required, on average, $17$ seconds. 
    \item[vi) ] \textit{Bayesian} prediction intervals for $\theta$ are computed using the {\tt R} package {\tt brms} and the function {\tt posterior\_epred}; see \cite{brms}.  We use a normal distribution prior with mean zero and standard deviation $4$ for $\mu$, and independent half-Cauchy prior distributions with scale parameter equal to 1 for the variance components.  We also used package rstanarm, which makes default, weakly-informative choices of prior distributions, and found this did not substantially affect the simulation results. Average run time was 3 seconds. 
\end{itemize}

In addition to the above methods, we evaluated a conformal prediction method \citep[see, e.g., ][]{cella.martin.2020} and two methods based on Satterthwaite approximations.  Because these methods did not perform well we did not include the corresponding results here.  However, the Appendix~\ref{A:satt}---\ref{A:sims} include detailed descriptions of these methods as well as additional simulations results.      

Table~1 provides results of our simulation study for predicting a new group mean $\theta$.  The nominal coverage of all intervals displayed in Table~2 is $95\%$, except for the adjustment to the conservative joint IM intervals, which have nominal level $90\%$.  Besides the $95\%$ intervals summarized in Table~1 we compared prediction intervals over a wide range of coverage levels, and found similar patterns of under-, over-, and correct coverage.  We would like to highlight three main take-away messages from our simulation study:
\begin{enumerate}
    \item As claimed, the joint and generalized IM methods produce valid prediction intervals over all simulations and for any nominal coverage level.  While these methods are, predictably, somewhat conservative, at least in some cases, the heuristic adjustments we discussed in Section~\ref{SS:validity} improve their efficiency without sacrificing validity.  Only in the $(\sigma_\alpha^2, \sigma_\eps^2) = (1.0, 0.1)$ scenarios did the Student's $t$ intervals attain their nominal coverage.  And, in those cases, the generalized IM intervals were just as efficient.  Compared to the Student's $t$ and parametric bootstrap intervals, which slightly under-cover in most cases, the generalized and adjusted generalized IM intervals are just enough wider to attain nominal coverage, and not overly conservative.  
    \item When the between-group variance is close (but not too close) to zero, i.e., when $(\sigma_\alpha^2, \sigma_\eps^2) = (0.1, 1.0)$ its restricted maximum likelihood estimate is often very close to zero. For example, in setting B about $86\%$ of simulated MLEs $\hat\sigma_a^2$ were less than $0.0001$.  The Student's $t$ prediction intervals simply plug-in the point estimates for the variance components, and, as a result, tend to undercover substantially in this case.  The performance of Student's $t-$based intervals did not necessarily improve with increased sample size; compare settings A to B and C to D for $(\sigma_a^2, \, \sigma^2) = (0.1,\,1.0)$.  The choice of $I-2$ degrees of freedom seems to be very conservative, and yet this method still experiences some under-coverage.  That suggests other Student's $t$-based prediction intervals, like those based on a Satterthwaite approximation, will not always attain nominal coverage; and, see the additional simulation results available in Appendix~\ref{A:sims} which show that, indeed, such intervals do under-cover.  
    \item Both the bootstrap and Bayesian alternatives suffered under-coverage for every pair of variance component values.  The low average lengths of these intervals, in some cases shorter than the oracle intervals, suggests these methods systematically produce intervals that are too short.   
\end{enumerate}

\begin{table}[h]
\caption{Observed coverage proportion and ratios of average prediction interval lengths compared to the Oracle method of $95\%$ prediction intervals for $\theta$. Bold text denotes significant under--coverage ($<93.5\%$ coverage).}
\centering
\resizebox{1\textwidth}{!}{%
\begin{tabular}{@{}cccccccccc@{}}\hline
                            &               & \multicolumn{8}{c}{Simulation Setting}                                                      \\\cmidrule(l){3-10} 
                            &               & \multicolumn{2}{c}{A} & \multicolumn{2}{c}{B} & \multicolumn{2}{c}{C} & \multicolumn{2}{c}{D} \\
$(\sigma_\alpha^2, \, \sigma_\eps^2)$ & Method        & Coverage   & Length   & Coverage   & Length   & Coverage   & Length   & Coverage   & Length   \\ \hline
$(0.1, \,1.0)$              & Oracle        & 0.94       & ---     & 0.95       & ---     & 0.94       & ---     & 0.95       & ---    \\
& Student $t$   & \hlgray{0.92}      & 1.35     & \hlgray{0.86}      & 1.00     & \hlgray{0.91}       & 1.28     & \hlgray{0.85}      & 0.99     \\
                            & Joint IM           & 0.98      &    2.64  &   0.98     &    1.71  & 0.99       & 3.15     &    0.98    &1.93     \\
                            & Adj. Joint IM      & 0.96 & 2.01 & 0.97  & 1.40    &0.97      & 2.30  & 0.97  & 1.54      \\
                            & Gen. IM            &  0.98     &  1.96   &  0.98      & 1.47   &  0.99   & 2.07    &   0.99    & 1.59      \\
                            & Adj. Gen. IM      &   0.96    & 1.69     &   0.94    & 1.23    & 0.97    & 1.76    &    0.95   & 1.25  \\ 
                            & Nonpar. Boot. & 0.99       & 1.44     & 0.99       & 1.38     & 0.98       & 1.50     & 0.99       & 1.53     \\
                            & Para. Boot.   & 0.95       & 1.33     & \hlgray{0.92}   & 1.08     & 0.95       & 1.33     & \hlgray{0.93}       & 1.09     \\
                            & Bayes         & 0.94       & 1.11     & \hlgray{0.90}   & 0.96     & 0.95       & 1.13     & \hlgray{0.91}      & 0.98     \\\hline
$(0.5, \,0.5)$              & Oracle        & 0.94       & ---     & 0.95       & ---     & 0.94       & ---     & 0.94       & ---     \\
                    & Student $t$   & \hlgray{0.91}   & 1.34     & \hlgray{0.93}       & 1.08     & \hlgray{0.90}      & 1.31     & 0.94       & 1.08     \\
                            & Joint IM           &   0.98    &    2.05  &    0.98    &    1.50  & 0.98       & 2.27      &     0.99   & 1.64     \\
                            & Adj. Joint IM     & 0.96  & 1.59 & 0.98 & 1.26 & 0.96 & 1.72  & 0.97 & 1.34 \\
                            & Gen. IM            &   0.96    & 1.45    &   0.95  &  1.17    &   0.96   &    1.46  &    0.95   & 1.19     \\
                            & Adj. Gen. IM       &    0.94   & 1.43     &   0.95    & 1.14   &  0.94   &1.45     &   0.94    & 1.14  \\ 
                    & Nonpar. Boot. & \hlgray{0.82}   & 0.80     & \hlgray{0.89}      & 0.88     & \hlgray{0.85}      & 0.80     & \hlgray{0.90}       & 0.90    \\
                     & Para. Boot.   & \hlgray{0.88}      & 1.07     & \hlgray{0.93}       & 1.03     & \hlgray{0.89}      & 1.06     & \hlgray{0.92}      &1.03     \\
                      & Bayes         & \hlgray{0.79}      & 0.74     & \hlgray{0.86}      & 0.83     & \hlgray{0.81}       & 0.73     & \hlgray{0.87}       & 0.83     \\\hline
$(1.0, \,0.1)$              & Oracle        & 0.95       & ---     & 0.94       & ---     & 0.94       & ---     & 0.94       & ---     \\
                            & Student $t$   & 0.95       & 1.38     & 0.94       & 1.09     & 0.95       & 1.37     & 0.94       &1.09     \\
                            & Joint IM          &  0.98      & 1.92     &   0.98     & 1.47     &   0.98     & 1.97     &   0.98     &  1.52    \\
                            & Adj. Joint IM     &  0.96 & 1.51  & 0.97 & 1.24  & 0.97 & 1.55  &  0.97 & 1.27   \\
                            & Gen. IM            &  0.95   & 1.36      &  0.94     &     1.13&  0.95      & 1.36    &  0.94    &    1.13  \\
                            & Adj. Gen. IM       &    0.95   & 1.36     &    0.94   & 1.13    &   0.95  & 1.36     &    0.94   & 1.13  \\ 
                            & Nonpar. Boot. & \hlgray{0.72}   & 0.61     & \hlgray{0.84}      & 0.78     & \hlgray{0.72}      & 0.61     & \hlgray{0.84}       & 0.78     \\
                            & Para. Boot.   & \hlgray{0.89}   & 1.05     & \hlgray{0.93}       &  1.03    & \hlgray{0.90}       & 1.05     & \hlgray{0.93}       & 1.03     \\
                            & Bayes         & \hlgray{0.72}   & 0.61     & \hlgray{0.84}  & 0.78     & \hlgray{0.72}   & 0.61     & \hlgray{0.84}      & 0.77    \\\hline
\end{tabular}%
}
\end{table}

\ifthenelse{1=1}{}{
\begin{table}[h]
\caption{Observed coverage proportion and ratios of average prediction interval lengths compared to the Oracle method of $95\%$ prediction intervals for $\theta$. Gray highlighting denotes significant under--coverage ($<93.5\%$ coverage).}
\centering
\resizebox{1\textwidth}{!}{%
\begin{tabular}{@{}cccccccccc@{}}\hline
                            &               & \multicolumn{8}{c}{Simulation Setting}                                                      \\\cmidrule(l){3-10} 
                            &               & \multicolumn{2}{c}{A} & \multicolumn{2}{c}{B} & \multicolumn{2}{c}{C} & \multicolumn{2}{c}{D} \\
$(\sigma_a^2, \, \sigma^2)$ & Method        & Coverage   & Length   & Coverage   & Length   & Coverage   & Length   & Coverage   & Length   \\ \hline
$(0.01, \,1.0)$             & Oracle        & 0.94       & ---     & 0.96       & ---     & 0.94       & ---     & 0.96       & ---    \\
                            & Student $t$   & 0.99       & 1.83     & 0.94       & 1.20     & 0.99       & 1.80     & 0.94       & 1.20     \\
                            & IM            & 0.98       & 2.50    & 0.97       & 2.10     & 1.00       & 2.54     & 1.00       & 2.09     \\
                            & Nonpar. Boot. & 1.00       & 2.41     & 1.00       & 2.82     & 1.00       & 2.57     & 1.00       & 3.27     \\
                            & Para. Boot.   & 1.00       & 2.00    & 1.00       & 1.67     & 1.00       & 2.05     & 1.00       & 1.69     \\
                            & Bayes         & 1.00       & 1.76     & 0.99       & 1.56     & 1.00       & 1.84     & 1.00       & 1.62     \\\hline
$(0.1, \,1.0)$              & Oracle        & 0.94       & ---     & 0.95       & ---     & 0.94       & ---     & 0.95       & ---    \\
& Student $t$   & \hlgray{0.92}      & 1.35     & \hlgray{0.86}      & 1.00     & \hlgray{0.91}       & 1.28     & \hlgray{0.85}      & 0.99     \\
                            & IM            & 0.95       & 1.72     & 0.94       & 1.40     & 0.98       & 1.61     & 0.96       & 1.28     \\
                            & Nonpar. Boot. & 0.99       & 1.44     & 0.99       & 1.38     & 0.98       & 1.50     & 0.99       & 1.53     \\
                            & Para. Boot.   & 0.95       & 1.33     & \hlgray{0.92}   & 1.08     & 0.95       & 1.33     & \hlgray{0.93}       & 1.09     \\
                            & Bayes         & 0.94       & 1.11     & \hlgray{0.90}   & 0.96     & 0.95       & 1.13     & \hlgray{0.91}      & 0.98     \\\hline
$(0.5, \,0.5)$              & Oracle        & 0.94       & ---     & 0.95       & ---     & 0.94       & ---     & 0.94       & ---     \\
                    & Student $t$   & \hlgray{0.91}   & 1.34     & \hlgray{0.93}       & 1.08     & \hlgray{0.90}      & 1.31     & 0.94       & 1.08     \\
                            & IM            & 0.94       & 1.53     & 0.97       & 1.26     & 0.94       & 1.44     & 0.94       & 1.14     \\
                    & Nonpar. Boot. & \hlgray{0.82}   & 0.80     & \hlgray{0.89}      & 0.88     & \hlgray{0.85}      & 0.80     & \hlgray{0.90}       & 0.90    \\
                     & Para. Boot.   & \hlgray{0.88}      & 1.07     & \hlgray{0.93}       & 1.03     & \hlgray{0.89}      & 1.06     & \hlgray{0.92}      &1.03     \\
                      & Bayes         & \hlgray{0.79}      & 0.74     & \hlgray{0.86}      & 0.83     & \hlgray{0.81}       & 0.73     & \hlgray{0.87}       & 0.83     \\\hline
$(1.0, \,0.1)$              & Oracle        & 0.95       & ---     & 0.94       & ---     & 0.94       & ---     & 0.94       & ---     \\
                            & Student $t$   & 0.95       & 1.38     & 0.94       & 1.09     & 0.95       & 1.37     & 0.94       &1.09     \\
                            & IM            & 0.96       & 1.47     & 0.95       & 1.14     & 0.95       & 1.36     & 0.95       & 1.15     \\
                            & Nonpar. Boot. & \hlgray{0.72}   & 0.61     & \hlgray{0.84}      & 0.78     & \hlgray{0.72}      & 0.61     & \hlgray{0.84}       & 0.78     \\
                            & Para. Boot.   & \hlgray{0.89}   & 1.05     & \hlgray{0.93}       &  1.03    & \hlgray{0.90}       & 1.05     & \hlgray{0.93}       & 1.03     \\
                            & Bayes         & \hlgray{0.72}   & 0.61     & \hlgray{0.84}  & 0.78     & \hlgray{0.72}   & 0.61     & \hlgray{0.84}      & 0.77    \\\hline
\end{tabular}%
}
\end{table}}

\section{Applications}
\label{S:examples}
\subsection{Soybean yield and fungicide use in Iowa}
\label{SS:soydata}
In this section we analyze soybean yields from $37$ Iowa farms comparing the effect of Stratego fungicide use on yield versus current growing practices that omit fungicide.  To model this data we use the random intercept model $Y_{ij} = \mu + \alpha_i + \eps_{ij}$ where $Y_{ij}$ denotes the natural logarithm of yield proportions (log of the response ratio) for strip pair $j$ on farm $i$.  Treated and non-treated strips are paired so that the response $Y_{ij}$ is itself an observation of the treatment effect.  The parameter $\mu$ denotes the overall population-averaged treatment effect, $\alpha_i\stackrel{iid}{\sim}\mathsf{N}(0, \sigma_\alpha^2)$ is a random intercept term for the farm effect, and $\eps_{ij}\sim \mathsf{N}(0, \sigma_\eps^2)$ is the random sampling effect.  The experimental data is unbalanced, with farms using fungicide on between 3 and 12 strips, and contains a total of $200$ responses; and see \cite{laurent.2020}.   

\begin{figure}[h]
    \caption{Responses and mean responses over 37 Iowa farms.  Prediction intervals for a new farm mean response using six different methods are displayed at the bottom of the figure.  (A color version can be found in the electronic version of the article.)}
    \centering
    \includegraphics[width = 0.65\textwidth]{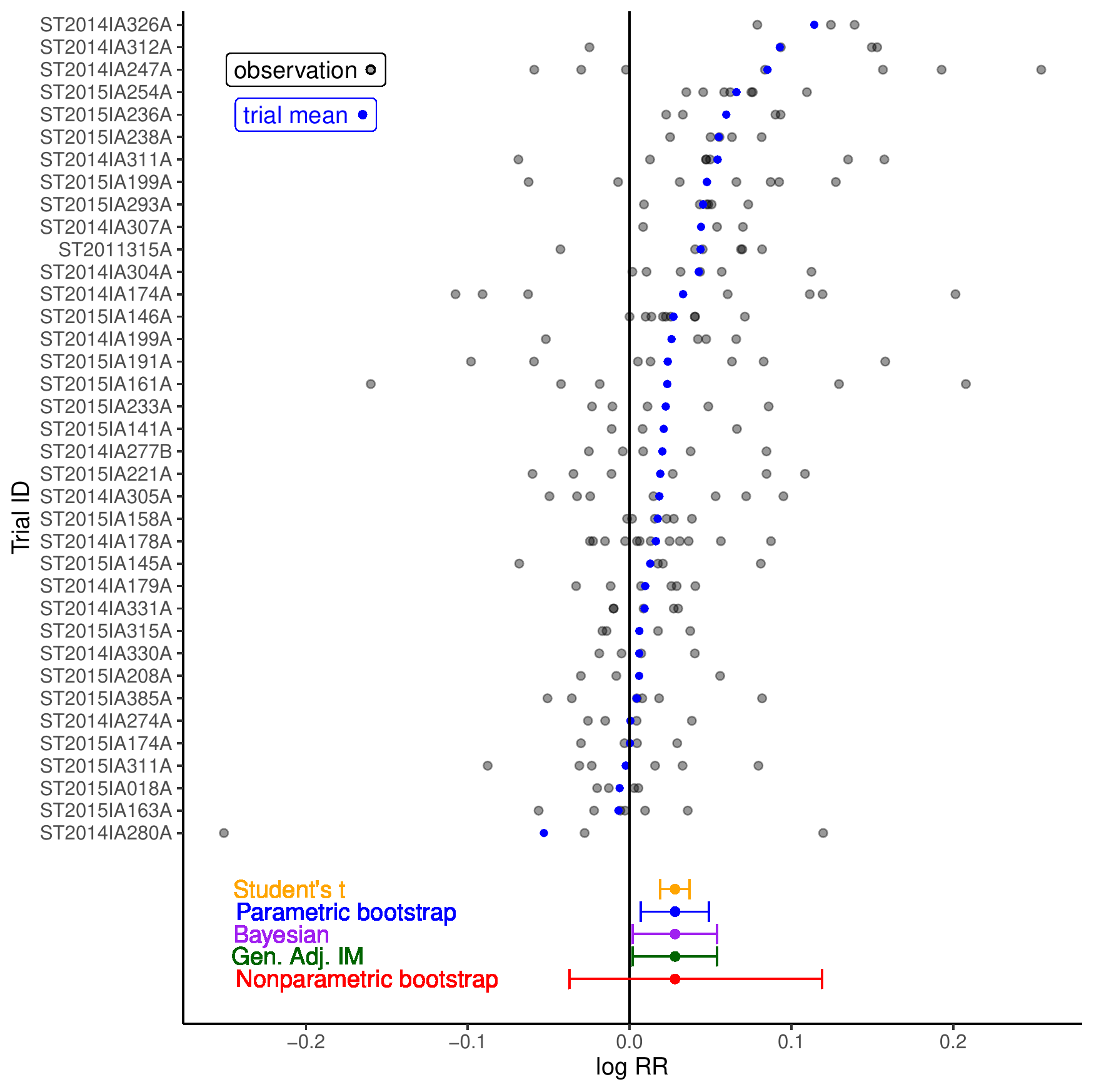}
    \label{fig:soydata}
\end{figure}

Figure~\ref{fig:soydata} displays ranges of fungicide effects across the farms and provides some sense of the relative magnitudes of between- and within-farm variance.  Within-farm variance is larger than between-farm variance, but it may be surprising that the restricted maximum likelihood estimate of the between-farm variance is zero.  Compared to the simulations in Section~\ref{S:simulations} this data set is most similar to setting D, which is a moderate sized, unbalanced experiment, and with variance component values of $(0.1,1.0)$, since, for this on-farm trial the between-farm variance estimate is zero.       


Given these similarities, we would expect the Student's $t$ prediction interval for a new farm mean response $\theta$ may be too short, since that method under-covered in that particular simulation.  Table~\ref{tbl:soysff} shows the prediction intervals for $\theta$ and $Y^\star$ for several methods, and, indeed, the Student's $t$ interval for $\theta$ is an outlier, being, by far, the shortest.  Based on those simulations we expect the joint IM and non-parametric bootstrap to produce conservative intervals, and recommend the adjusted generalized IM intervals as providing the best efficiency while still demonstrating validity across those simulations.  For the on-farm trial the adjusted generalized IM, Bayesian, and parametric bootstrap intervals are almost indistinguishable.  They all suggest a small, positive fungicide effect for a new farm mean, while predicting new observations of strip pairs are likely to show no effect, or even a negative effect.


\ifthenelse{1=1}{}{\begin{figure}[h]
    \caption{Fused plausibility contour $\pi_n^f$ (black curve) for a new farm mean fungicide effect $\theta$ for the data in Section~\ref{SS:soydata}, along with $\eta-$specific contours $\pi_n^{\eta}$ (grey fill) over grid of $\eta$ values.}
    \label{fig:soysff_plaus}
    \centering
    \includegraphics[width = .6\columnwidth]{soysff_curve_annotated.pdf}
\end{figure}}

\begin{table}[h]
\centering
\caption{$95\%$ prediction intervals for $\theta$ and a single new response $Y^\star$ on soybean yield for the data described in Section~\ref{SS:soydata}.\label{tbl:soysff}}
\begin{tabular}{@{}lcc@{}}\toprule
                         & \multicolumn{2}{c}{$95\%$ Prediction Intervals}                                 \\ \cmidrule(l){2-3} 
Method                   & \multicolumn{1}{c}{$\theta$} & \multicolumn{1}{c}{$Y^\star$} \\ \midrule
Student $t$              &       $(0.019, \, 0.037)$                         &               $(-0.096, \,  0.152)$                             \\
Para. Boot.     &       $(0.007,\,  0.049)$                             & $(-0.094, \,  0.150)$                                            \\
Bayesian & $(0.005, 0.055)$ & $(-0.098, 0.152)$ \\
Joint IM                       &    $(-0.014,\,  0.070)$                                 &     $(-0.132, \,  0.188)$   \\
Adj. Joint IM  &    $(-0.006,\,  0.062)$                                 &     $(-0.111, \,  0.167)$   \\
Gen. IM  &    $(-0.037,\, 0.092)$                                 &     $(-0.107, \,  0.163)$   \\
Adj. Gen. IM  &    $(0.002,\,  0.054)$                                 &     $(-0.107, \,  0.163)$   \\
Nonpar. Boot. &   $(-0.037,\,  0.119)$                                 &      $(-0.082, \, 0.162)$                                      \\
 \bottomrule
\end{tabular}
\end{table}

\subsection{Livestock diets and average daily weight gain}
\label{SS:example2}

In this section we analyze a benchmark data set for mixed effects models included in \citet{SASbook} as data set 5.3.  The data comes from a designed experiment to examine the effects of four diets including different levels of medication ($0, 10, 20,$ or $30$) on the average daily weight gain of steers.  The experimenters controlled for initial weight at the start of the trial and also recorded the barns housing each steer---these contributed a random intercept to the model, which has the form
\[Y_{ij} = \beta_0 + \beta_1 x_{1,ij} + \beta_2 x_{2,ij} + \beta_3 x_{3,ij} + \beta_4 x_{4,ij} + \alpha_iz_{ij} + \eps_{ij},\]
where $Y_{ij}$ is the average daily weight gain of steer $j$ in barn $i$ over the course of the trial, $\beta_0$ is the intercept term which includes steers receiving treatment $0$, $\beta_1$ is the effect of initial weight $x_{1,ij}$, and $\beta_2, \beta_3$, and $\beta_4$ are the effects of diets with quantities $10$, $20$, and $30$ of medicine in relation to the baseline diet including no medicine; each barn includes one steer receiving each medicine amount.  Additionally, $z_{ij}$ records the barn housing each steer, $\alpha_i\stackrel{iid}{\sim} \mathsf{N}(0, \sigma_\alpha^2)$ is a random intercept term representing the variation in average daily weight gain over barns, and $\eps_{ij} \stackrel{iid}{\sim} \mathsf{N}(0, \sigma^2)$ represents random sampling variability.  The data contains $32$ responses over $8$ barns.     

\begin{figure}[h]
    \caption{Average daily weight gain for steers colored by barn, along with regression lines corresponding to each diet. (A color version can be found in the electronic version of the article.)}
    \label{fig:im_weightgain_regs}
    \centering
    \includegraphics[width = .6\columnwidth]{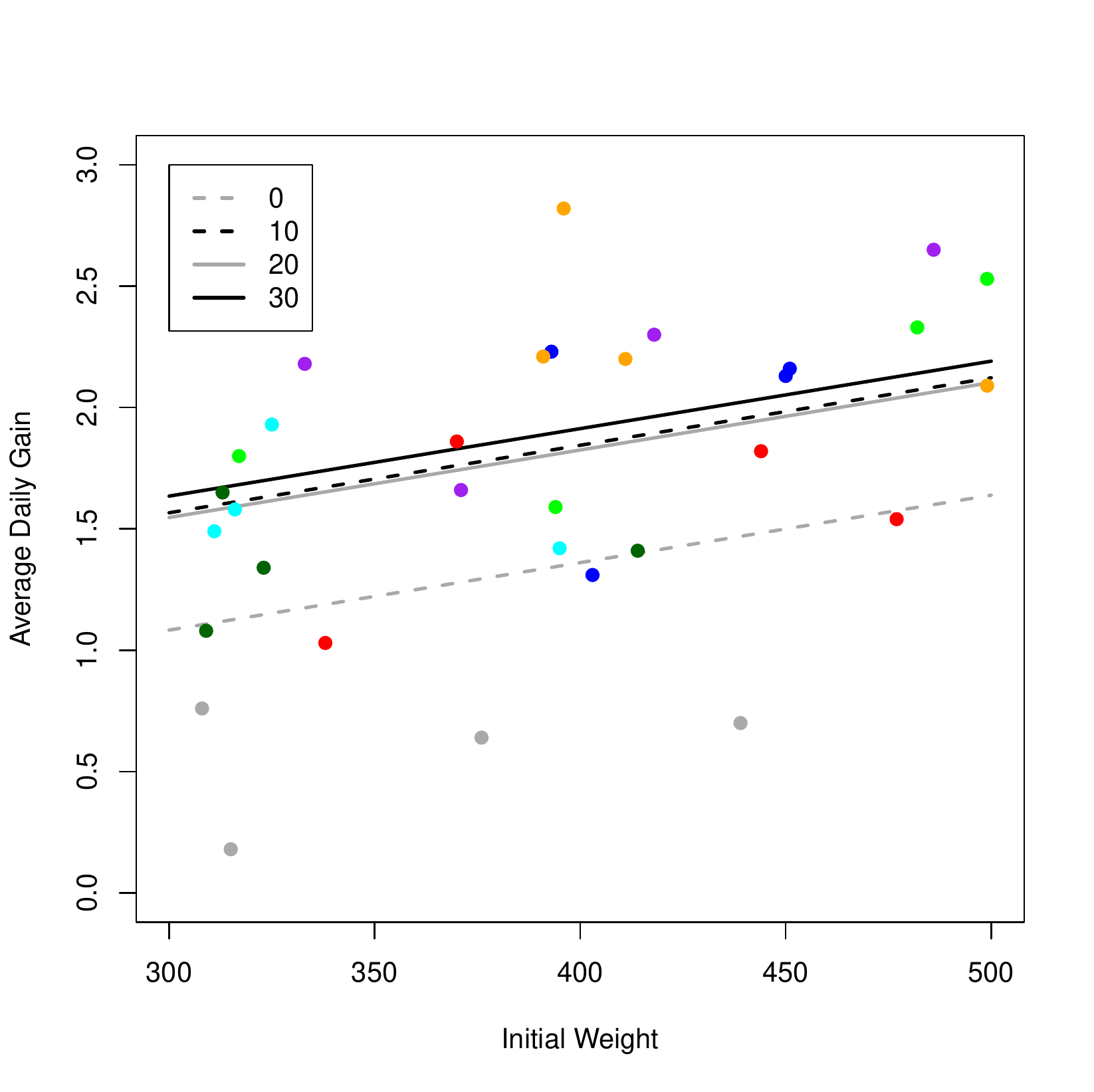}
\end{figure}

Figure~\ref{fig:im_weightgain_regs} displays responses grouped by barn along with regression lines for each treatment.  The plot illustrates substantial between-barn variability; for example, one barn has responses falling below every fitted regression line while another barn's responses fall mostly above every line.  The restricted maximum likelihood estimate of between-barn variance is about $0.24$, while the estimate of within-barn variance is only $0.05$.  That makes this data most similar to simulation setting A with variance component pair $(1.0,0.1)$.  In that simulation, the Bayesian and bootstrap methods under-covered while the generalized IM method was less conservative than the joint IM.      
Table~\ref{tbl:weight} displays $95\%$ prediction intervals of $\theta$ and $Y^\star$ for a new steer with an initial weight of 400 and treated with medication at level 10 using the IM, bootstrap, and Bayesian methods.  And, as expected based on the simulation setting A, the IM-based intervals are all wider than the Bayesian and bootstrap intervals.  Nevertheless, the adjusted, generalized IM intervals predict a positive diet effect on weight gain for a new barn mean as well as a new steer.  Given the under-coverage of Bayesian and bootstrap intervals in the simulation, we would recommend the IM intervals as more honest reflections of uncertainty.    
\begin{table}[h]
\centering
\caption{$95\%$ prediction intervals for $\theta$ and a single new response $Y^\star$ on average daily weight gain for the data described in Section~\ref{SS:example2}.\label{tbl:weight}}
\begin{tabular}{@{}lcc@{}}\toprule
                         & \multicolumn{2}{c}{$95\%$ Prediction Intervals}                                 \\ \cmidrule(l){2-3} 
Method                   & \multicolumn{1}{c}{$\theta$} & \multicolumn{1}{c}{$Y^\star$} \\ \midrule
Joint IM                         &    $(-0.11,\, 3.91)$                                 &     $(-0.19,\, 3.99)$   \\
Adj. Joint IM                    &    $(0.29,\, 3.51)$                                 &     $(0.22,\, 3.58)$   \\
Gen. IM                          &    $(0.47,\, 3.33)$                                 &     $(-0.73,\, 4.53)$   \\
Adj. Gen. IM                     &    $(0.50,\, 3.31)$                                 &     $(0.25,\, 3.55)$   \\
Para. Boot.                      &    $(0.74,\, 3.00)$                                 &     $(0.65,\, 3.05)$                                            \\
Bayesian                         &    $(0.76,\, 2.93)$                               &     $(0.64,\, 2.99)$ \\
\bottomrule
\end{tabular}
\end{table}

\ifthenelse{1=1}{}{
\begin{figure}
     \centering        \caption{One and two-dimensional IM plausibility contours for $\theta$ and $(\theta, \rho)$ compared to stacked prediction intervals using parametric bootstrap and Bayesian methods.  Horizontal cuts of one-dimensional contours at height $\alpha$ identify $100(1-\alpha)\%$ prediction intervals for $\theta$.}
     \begin{subfigure}[b]{0.48\textwidth}
         \centering
         \includegraphics[width=\textwidth]{im_weightgain_plaus_comp.pdf}
         \caption{One-dimensional $\theta$ contours}
         \label{fig:theta_weight}
     \end{subfigure}
     \hfill
     \begin{subfigure}[b]{0.48\textwidth}
         \centering
         \includegraphics[width=\textwidth]{im_weightgain_plaus_joint2.pdf}
         \caption{Two-dimensional Joint IM contour}
         \label{fig:Y_weight}
     \end{subfigure}
     \hfill

        \label{fig:weight}
\end{figure}}

\section{Discussion}
\label{s:conclusion}

In this manuscript we applied the IM framework to prediction in two-stage linear mixed effects models.  Current methods do not produce valid prediction intervals for random effects associated with observations of new groups or individuals.  The IM method, on the other hand, provides provably valid predictions, which we also demonstrated in simulation experiments.  The simulation provided some justification for adjustments to the IM methods that improve their efficiency without sacrificing validity, and also provided intuition helpful for two real data analyses.   

The standard IM construction applied to the two-stage model resulted in conservative prediction intervals.  This may appear to be a downside of the IM framework, but it is really a reflection of the challenges to inference and prediction posed by nuisance parameters.  In prediction problems like the two-stage model, full marginalization of nuisance parameters is not possible.  In contrast to the IM framework, a typical frequentist strategy is to define an asymptotic-pivot---a function of the parameter of interest, data, and a consistent point estimator of the nuisance parameter that has a sampling distribution convergent (as $n\rightarrow \infty$) to one that depends on no unknowns.  These plug-in estimation methods sacrifice (at least finite-sample) validity for efficiency.  The IM mindset is to insist on validity, but the heuristic adjustments we make suggest a compromise strategy is valuable.

Practitioners often rely on large-sample results to justify the use of plug-in methods, like the Student's $t$ prediction intervals, or the bootstrap.  For simple, one-sample problems, these asymptotic results ``kick in" quickly. For mixed models it is less clear what sample size is needed in order for inferences and predictions based on large-sample results to be reliable.  It seems likely that practitioners using such methods for linear mixed models place too much faith in their predictions in small and moderate sized experiments.  IM methods for mixed models should be developed further to provide valid predictions in such applications.  And, better computational tools, like {\tt R} packages, are needed to improve usability of IM methods for practitioners.

\section*{Acknowledgments}

This article is a product of the Iowa Agriculture and Home Economics Experiment Station, Ames, Iowa. Project No. IOW03717 is supported by USDA/NIFA and State of Iowa funds.  Any opinions, findings, conclusions, or recommendations expressed in this publication are those of the author(s) and do not necessarily reflect the views of the U.S. Department of Agriculture.

\appendix
\section{Technical Details from Section 4}
\subsection{Details for the association in (5)}
\label{A:assoc.suff}
These details are reproduced from \citet{martin.liu.book} Section 8.3 for completeness.  From \eqref{eq:model}, make the one-to-one transformation $Y\mapsto (K^\top Y, \, BY)$ where $K$ is an $n\times (n-p)$ matrix such that $KK^\top = I_n - X(X^\top X)^{-1}X^\top$ and $K^\top K = I_{n-p}$, and where $B = (X^\top X)^{-1}X^\top$.  Then, 
\begin{equation}
\label{eq:BY}
\begin{aligned}
    &K^\top Y\sim \mathsf{N}_{n-p}(0,\sigma_\eps^2 I_{n-p} + \sigma_\alpha^2 H), \,\text{and}
		&BY \sim \mathsf{N}_{p}(\beta, C_\sigma),
\end{aligned}
\end{equation}
where $H = K^\top G K$ and $C_\sigma = (\sigma_\eps^2 BB^\top + \sigma_\alpha^2 B G B^\top)$.  

Let $P$ diagonalize $H$ such that $P^\top H P = \lambda I_{n-p}$ is equal to the identity matrix multiplied by the $(n-1)\times 1$ vector of eigenvalues of $H$, denoted $\lambda$.  $P$ may be written $P = [P_1, \ldots, P_L]$ where $L$ is the number of distinct eigenvalues of $G$ and $P_\ell$ is an $(n-p)\times r_\ell$ matrix where $r_\ell$ is the multiplicity of $\lambda_\ell$.  Define $S_\ell = Y^\top KP_\ell P_\ell^\top K^\top Y$.  Then, $(S_1,\ldots, S_L)$ are minimal sufficient for $(\sigma_\alpha^2, \sigma_\eps^2)$ and 
\begin{equation}
\label{eq:SL}
\begin{aligned}
    &S_\ell = (\lambda_\ell \sigma_\alpha^2 + \sigma_\eps^2)V_\ell,\quad V_\ell\stackrel{ind.}{\sim} \chi^2(r_\ell), \quad \ell = 1,\ldots, L.
    \end{aligned}
\end{equation}

\subsection{Details for the association in (8)}
\label{A:local.cond}

See also \citet{martin.hc}.  Let $g_\ell(\rho) = \frac{\partial}{\partial \rho}\frac{1+\rho(\lambda_\ell - 1)}{1+\rho(\lambda_L - 1)}$ for $\ell = 1, \ldots, L-1$, and let $g(\rho) = (g_1(\rho), \ldots, g_{L-1}(\rho))^\top$.  Define the matrix $M_0$ on which (8) depends to be any $(L-1)\times (L-2)$ matrix orthogonal to $g(\rho_0)$ for a particular value $\rho_0$.    

Let $\tau = \left(\log \left[(V_1 / V_L)(r_L/r_1)\right], \ldots, \log \left[(V_{L-1} / V_{L})(r_{L}/r_{L-1})\right]\right)^\top M_0$ and let $H$ be equal to the observed value 
\[\left(\log\left[\frac{S_1}{S_L}\frac{r_L}{r_1}\right] - \log\left[\frac{\rho_0(\lambda_1-1)+1}{\rho_0(\lambda_L-1)+1}\right], \ldots, \log\left[\frac{S_{L-1}}{S_L}\frac{r_L}{r_{L-1}}\right] - \log\left[\frac{\rho_0(\lambda_{L-1}-1)+1}{\rho_0(\lambda_L-1)+1}\right]\right)^\top M_0.\]
Let $M_0'$ be the matrix formed by prepending the column vector $(1,0_{L-2})^\top$ to $M_0$; $M_0'$ has full rank.  Let $u$ be a scalar and $(u')^\top = (u, H){M_0'}^{-1}$ be an $(L-1)\times 1$ vector.

Then, the joint density of $(U_0, WV_L^{-1/2}|\tau = H)$ on the log scale and up to an additive constant is given by
\[f(u,v) = \tfrac12\sum_{\ell = 1}^{L-1}r_\ell u'_\ell  - \tfrac12\left(1+\sum_{\ell=1}^L r_\ell\right)\log\left(1/2+\tfrac{v^2}{2r_L} + \tfrac{1}{2r_L}\sum_{\ell = 1}^{L-1}r_\ell e^{u'_\ell}\right).\]

\section{Satterthwaite approximations}
\label{A:satt}
According to \eqref{eq:assoc.unreduced} and the details regarding that association presented in Appendix~\ref{A:assoc.suff} we have
\[S_\ell = (\lambda_\ell\sigma_\alpha^2 + \sigma_\eps^2)V_\ell, \quad V_\ell \stackrel{ind.}{\sim}\chi^2(r_\ell), \quad \ell = 1, \ldots, L.\]
The goal is to set $c, \,\nu >0$ such that
\[\frac{(\theta - x^\top \hat\beta)}{\sqrt{\frac{c\sum_{\ell = 1}^L S_\ell}{\nu}}}\stackrel{\cdot}{\sim}t(\nu).\]
Let $S':=\frac{E(\sum_{\ell = 1}^L S_\ell)}{\tfrac12 V(\sum_{\ell = 1}^L S_\ell)}\sum_{\ell = 1}^L S_\ell$ so that $V(S') = 2E(S')$ by construction.  The $\chi^2$ distribution with first two moments matching those of $S'$ has degrees of freedom
\[\nu = \frac{[\sum_{\ell = 1}^L r_\ell(\lambda_\ell\sigma_\alpha^2+\sigma_\eps^2)]^2}{\sum_{\ell = 1}^L r_\ell(\lambda_\ell\sigma_\alpha^2+\sigma_\eps^2)^2}.\]
Simplify the fraction $\frac{E(\sum_{\ell = 1}^L S_\ell)}{\tfrac12 V(\sum_{\ell = 1}^L S_\ell)} \cdot \nu^{-1}$ to see that
\[\frac{(\theta-x^\top\hat\beta)/(c_1\sigma_\alpha^2 + c_2\sigma_\eps^2)}{\sqrt{\frac{\sum_{\ell = 1}^L S_\ell}{\sum_{\ell = 1}^L r_\ell(\lambda_\ell\sigma_\alpha^2+\sigma_\eps^2)}}}\sim t(\nu)\]
for $(c_1, c_2)$ as defined in Section~\ref{ss:associate}.  Replacing $(\sigma_\alpha^2, \sigma_\eps^2)$ with their respective restricted maximum likelihood estimates yields an approximate Student's $t$ pivot for $\theta$.  We can obtain approximate pivots for $\theta$ or a new observation $Y^\star$ by making the appropriate modifications to the constants $(c_1, c_2)$.  Inverting the approximate pivot yields an approximate prediction interval for $\theta$ or $Y^\star$.    

This is not the only way to construct prediction intervals based on an approximate pivot with a Student's $t$ distribution.  \citet{Francq.etal.2019} use a generalized Satterthwaite method to determine the degrees of freedom $\tau$ used in the following interval:
\[x^\top \hat\beta \pm t_{1-\alpha/2,\, \tau} \sqrt{c_1\hat\sigma_\alpha^2 + c_2\hat\sigma_\eps^2}\]
where $(\hat\sigma_\alpha^2, \hat\sigma_\eps^2)$ are restricted maximum likelihood estimates of the variance components.  This interval has the same form, but with a different choice of degrees of freedom, as the Student's $t$ interval used in the simulations in Section~\ref{S:simulations}.

\section{Further Simulation Results}
\label{A:sims}

In addition to the simulation results reported in Section~5 we also evaluated the performance of those methods  for predicting new responses; see Table~\ref{table:new} below.  Similar to the simulations for predicting a new group mean, the IM method consistently attains or exceeds its nominal coverage level.  The Student $t$ intervals perform better with respect to coverage level for new responses compared to a new group mean, but are less efficient than the IM intervals. Again, the bootstrap and Bayesian prediction intervals often fail to cover at the nominal level when predicting a response from a new group, but fare better at predicting a new response from an existing group.     

Tables~\ref{table:theta_satt} and \ref{table:new_satt} display the results of the same simulations using the Satterthwaite and generalized Satterthwaite methods described in Appendix~\ref{A:satt}.  Both methods perform well for predicting a new response, similar to the performance of the IM.  However, they both experience substantial under-coverage when predicting a new group mean.  The degrees of freedom selected by the generalized Satterthwaite method tends to be larger than $I-2$, the suggested degrees of freedom according to  \cite{higgins.etal.2009}, making those intervals shorter and, hence, tending to cover less often.  It is not obvious how to expect the Satterthwaite method to perform by comparison due to its different construction using the association. The simulations show that it often produces intervals slightly longer and with slightly better coverage than the generalized Satterthwaite method, but still experiences worse coverage performance than the intervals suggested by \cite{higgins.etal.2009}.

\begin{table}
\caption{Observed coverage proportion and ratios of average prediction interval lengths compared to the Oracle method of $95\%$ prediction intervals for a new observation $Y^\star$ in a new group. Gray highlighting denotes significant under--coverage.}
\label{table:new}
\centering
\resizebox{1\textwidth}{!}{%
\begin{tabular}{@{}cccccccccc@{}}\hline
                            &               & \multicolumn{8}{c}{Simulation Setting}                                                        \\\cmidrule(l){3-10} 
                            &               & \multicolumn{2}{c}{A} & \multicolumn{2}{c}{B} & \multicolumn{2}{c}{C} & \multicolumn{2}{c}{D} \\
$(\sigma_a^2, \, \sigma^2)$ & Method        & Coverage   & Length   & Coverage   & Length   & Coverage   & Length   & Coverage   & Length   \\ \hline
$(0.1, \,1.0)$              & Oracle        & 0.95       & ---     & 0.96       & ---     & 0.96       & ---     & 0.96       & ---     \\
                            & Student $t$   & 1.00       & 1.57     & 0.98       & 1.17     & 1.00       & 1.57     & 0.98       & 1.17     \\
                            & Joint IM           &    0.98    & 1.37     &  0.99      & 1.27     &    0.99    & 1.54     &    0.99    &   1.34   \\
                            & Adj. Joint IM           &  0.96      &1.15      &  1.97      &    1.11  &    0.97    &    1.25  &   0.98     &    1.16  \\
                            & Gen. IM           &   0.97     & 1.62     &  0.99      & 1.59     &   0.98     & 1.58     &    0.99    &  1.58    \\
                            & Adj. Gen. IM           &   0.98     & 1.90     &   0.99     & 1.43     &    0.99    & 1.93     &    0.99    & 1.44     \\
                            & Nonpar. Boot. & \hlgray{0.92}       & 1.00     & 0.94       & 0.98     & \hlgray{0.92}       & 0.99     & 0.94       & 0.98     \\
                            & Para. Boot.   & 0.95       & 1.02     & 0.95       & 1.00     & 0.95       & 1.02     & 0.96       & 1.00     \\
                            & Bayes         & 0.96       & 1.05     & 0.96       & 1.01     & 0.96       & 1.06     & 0.96       & 1.01     \\\hline
                            
$(0.5, \,0.5)$              & Oracle        & 0.95       & ---     & 0.95       & ---     & 0.95       & ---     & 0.95       & ---     \\
                            & Student $t$   & 0.99       & 1.50     & 0.97       & 1.14     & 0.99       & 1.49     & 0.97       & 1.14     \\
                            & Joint IM           &  0.98      & 1.66     &    0.99    & 1.35     &    0.99    & 1.84     &    0.99    & 1.46     \\
                            & Adj. Joint IM           &   0.96     &    1.34  & 0.97       &   1.16   &    0.98    & 1.44      &   0.98     &  1.23    \\
                            & Gen. IM           &    0.99    & 2.45     &   1.00     &  2.80    &   0.99     &  2.31    &  1.00      & 2.77     \\
                            & Adj. Gen. IM           &   0.99     & 2.11     &  0.99      & 1.37     &    0.99    &     2.05 &   0.99     & 1.37     \\
                            & Nonpar. Boot. & \hlgray{0.88}       & 0.90     & \hlgray{0.92}       & 0.93     & \hlgray{0.88}       & 0.88     & \hlgray{0.91}       & 0.93     \\
                            & Para. Boot.   & \hlgray{0.92}       & 1.04     & 0.94       & 1.01     & \hlgray{0.93}       & 1.03     & 0.94       & 1.01     \\
                            & Bayes         & \hlgray{0.91}       & 0.93     & \hlgray{0.92}       & 0.95     & \hlgray{0.92}       & 0.92     & \hlgray{0.93}       & 0.95     \\\hline
                            
$(1.0, \,0.1)$              & Oracle        & 0.94       & ---     & 0.94       & ---     & 0.94       & ---     & 0.94       & ---     \\
                            & Student $t$   & 0.96       & 1.41     & 0.94       & 1.10     & 0.97       & 1.40     & 0.94       & 1.10     \\
                            & Joint IM           &    0.98    &  1.87    &   0.98     & 1.45     &   0.99     & 1.92     &   0.99     & 1.50      \\
                            & Adj. Joint IM           &  0.96      &  1.51    &   0.96     & 1.22     &   0.97     &  1.51     &   0.97     &   1.25   \\
                            & Gen. IM           &   1.00     & 2.93     &   1.00     &  3.56    &  1.00      &  2.74    &   1.00     & 3.49     \\
                            & Adj. Gen. IM           &   0.98     & 1.57     &    0.96    &     1.17 &      0.98  & 1.54     &  0.96      &  1.17     \\
                            & Nonpar. Boot. & \hlgray{0.78}       & 0.71     & \hlgray{0.87}       & 0.83     & \hlgray{0.75}      & 0.69     & \hlgray{0.86}       & 0.82     \\
                            & Para. Boot.   & \hlgray{0.90}      & 1.05     & \hlgray{0.92}       & 1.02     & \hlgray{0.90}      & 1.04     & \hlgray{0.93}       & 1.02     \\
                            & Bayes         & \hlgray{0.78}      & 0.72     & \hlgray{0.88}      & 0.84     & \hlgray{0.78}       & 0.70     & \hlgray{0.87}       & 0.84 \\\hline   
\end{tabular}%
}
\end{table}

\ifthenelse{1=1}{}{
\begin{table}
\caption{Observed coverage proportion and ratios of average prediction interval lengths compared to the Oracle method of $95\%$ prediction intervals for a new observation $Y^\star$ in a new group. Gray highlighting denotes significant under--coverage.}
\label{table:new}
\centering
\resizebox{1\textwidth}{!}{%
\begin{tabular}{@{}cccccccccc@{}}\hline
                            &               & \multicolumn{8}{c}{Simulation Setting}                                                        \\\cmidrule(l){3-10} 
                            &               & \multicolumn{2}{c}{A} & \multicolumn{2}{c}{B} & \multicolumn{2}{c}{C} & \multicolumn{2}{c}{D} \\
$(\sigma_a^2, \, \sigma^2)$ & Method        & Coverage   & Length   & Coverage   & Length   & Coverage   & Length   & Coverage   & Length   \\ \hline
$(0.01, \,1.0)$             & Oracle        & 0.96       & ---     & 0.96       & ---     & 0.96       & ---     & 0.96       & ---     \\
                            & Student $t$   & 1.00       & 1.58     & 0.98       & 1.17     & 1.00       & 1.59     & 0.98       & 1.17     \\
                            & Joint IM           &        &      &        &      &        &      &        &      \\
                            & Adj. Joint IM           &        &      &        &      &        &      &        &      \\
                            & Gen. IM           &        &      &        &      &        &      &        &      \\
                            & Adj. Gen. IM           &        &      &        &      &        &      &        &      \\
                            & Nonpar. Boot. & 0.94       & 1.01     & 0.95       & 0.99     & 0.94       & 1.00     & 0.95       & 0.99     \\
                            & Para. Boot.   & 0.95       & 1.02     & 0.95       & 0.99     & 0.96       & 1.03     & 0.96       & 1.00     \\
                            & Bayes         & 0.96       & 1.07     & 0.96       & 1.02     & 0.96       & 1.08     & 0.96       & 1.02     \\\hline
                            
$(0.1, \,1.0)$              & Oracle        & 0.95       & ---     & 0.96       & ---     & 0.96       & ---     & 0.96       & ---     \\
                            & Student $t$   & 1.00       & 1.57     & 0.98       & 1.17     & 1.00       & 1.57     & 0.98       & 1.17     \\
                            & Joint IM           &        &      &        &      &        &      &        &      \\
                            & Adj. Joint IM           &        &      &        &      &        &      &        &      \\
                            & Gen. IM           &        &      &        &      &        &      &        &      \\
                            & Adj. Gen. IM           &        &      &        &      &        &      &        &      \\
                            & Nonpar. Boot. & \hlgray{0.92}       & 1.00     & 0.94       & 0.98     & \hlgray{0.92}       & 0.99     & 0.94       & 0.98     \\
                            & Para. Boot.   & 0.95       & 1.02     & 0.95       & 1.00     & 0.95       & 1.02     & 0.96       & 1.00     \\
                            & Bayes         & 0.96       & 1.05     & 0.96       & 1.01     & 0.96       & 1.06     & 0.96       & 1.01     \\\hline
                            
$(0.5, \,0.5)$              & Oracle        & 0.95       & ---     & 0.95       & ---     & 0.95       & ---     & 0.95       & ---     \\
                            & Student $t$   & 0.99       & 1.50     & 0.97       & 1.14     & 0.99       & 1.49     & 0.97       & 1.14     \\
                            & IM - J           &        &      &        &      &        &      &        &      \\
                            & IM            & 0.95       & 1.19     & 0.95       & 1.05     & 0.96       & 1.24     & 0.95       & 1.06     \\
                            & Nonpar. Boot. & \hlgray{0.88}       & 0.90     & \hlgray{0.92}       & 0.93     & \hlgray{0.88}       & 0.88     & \hlgray{0.91}       & 0.93     \\
                            & Para. Boot.   & \hlgray{0.92}       & 1.04     & 0.94       & 1.01     & \hlgray{0.93}       & 1.03     & 0.94       & 1.01     \\
                            & Bayes         & \hlgray{0.91}       & 0.93     & \hlgray{0.92}       & 0.95     & \hlgray{0.92}       & 0.92     & \hlgray{0.93}       & 0.95     \\\hline
                            
$(1.0, \,0.1)$              & Oracle        & 0.94       & ---     & 0.94       & ---     & 0.94       & ---     & 0.94       & ---     \\
                            & Student $t$   & 0.96       & 1.41     & 0.94       & 1.10     & 0.97       & 1.40     & 0.94       & 1.10     \\
                            & IM - J           &        &      &        &      &        &      &        &      \\
                            & IM            & 0.95       & 1.29     & 0.94       & 1.10     & 0.94       & 1.33     & 0.95       & 1.13     \\
                            & Nonpar. Boot. & \hlgray{0.78}       & 0.71     & \hlgray{0.87}       & 0.83     & \hlgray{0.75}      & 0.69     & \hlgray{0.86}       & 0.82     \\
                            & Para. Boot.   & \hlgray{0.90}      & 1.05     & \hlgray{0.92}       & 1.02     & \hlgray{0.90}      & 1.04     & \hlgray{0.93}       & 1.02     \\
                            & Bayes         & \hlgray{0.78}      & 0.72     & \hlgray{0.88}      & 0.84     & \hlgray{0.78}       & 0.70     & \hlgray{0.87}       & 0.84 \\\hline   
\end{tabular}%
}
\end{table}
}

\ifthenelse{1=1}{}{
\begin{table}
\centering
\caption{Observed coverage proportion and ratios of average prediction interval lengths compared to the Oracle method of $95\%$ prediction intervals for a new observation $Y^\star$ in an existing group.}
\label{table:exs}
\resizebox{1\textwidth}{!}{%
\begin{tabular}{@{}cccccccccc@{}}\hline
                            &             & \multicolumn{8}{c}{Simulation Setting}                                                        \\\cmidrule(l){3-10} 
                            &             & \multicolumn{2}{c}{A} & \multicolumn{2}{c}{B} & \multicolumn{2}{c}{C} & \multicolumn{2}{c}{D} \\
$(\sigma_a^2, \, \sigma^2)$ & Method      & Coverage   & Length   & Coverage   & Length   & Coverage   & Length   & Coverage   & Length   \\ \hline
$(0.01, \,1.0)$             & Oracle      & 0.94       & ---     & 0.95       & ---     & 0.94       & ---     & 0.95       & ---     \\
                            & Student $t$ & 1.00       & 1.59     & 0.97       & 1.17     & 1.00       & 1.59     & 0.97       & 1.17     \\
                            & IM          & 0.95       & 1.04     & 0.95       & 1.01     & 0.95       & 1.03     & 0.95       & 1.02     \\
                            & Para. Boot. & 0.94       & 1.00     & 0.94       & 1.00     & 0.94       & 0.99     & 0.94       & 0.99     \\
                            & Bayes       & 0.95       & 1.06     & 0.95       & 1.02     & 0.95       & 1.05     & 0.95       & 1.02     \\\hline
                            
$(0.1, \,1.0)$              & Oracle      & 0.95       & ---     & 0.94       & ---     & 0.95       & ---     & 0.95       & ---     \\
                            & Student $t$ & 0.99       & 1.60     & 0.97       & 1.18     & 1.00       & 1.62     & 0.97       & 1.18     \\
                            & IM          & 0.95       & 1.05     & 0.95       & 1.02     & 0.96       & 1.05     & 0.95       & 1.02     \\
                            & Para. Boot. & 0.95       & 1.01     & 0.95       & 1.00     & 0.94       & 1.00     & 0.94       & 1.00     \\
                            & Bayes       & 0.96       & 1.04     & 0.95       & 0.99     & 0.96       & 1.04     & 0.96       & 0.99     \\\hline
                            
$(0.5, \,0.5)$              & Oracle      & 0.95       & ---     & 0.96       & ---     & 0.95       & ---     & 0.96       & ---     \\
                            & Student $t$ & 1.00       & 1.66     & 0.98       & 1.19     & 1.00       & 1.85     & 0.98       & 1.22     \\
                            & IM          & 0.97       & 1.15     & 0.97       & 1.05     & 0.98       & 1.13     & 0.97       & 1.05     \\
                            & Para. Boot. & 0.96       & 1.05     & 0.96       & 1.02     & 0.94       & 1.06     & 0.96       & 1.02     \\
                            & Bayes       & 0.95       & 0.84     & 0.94       & 0.76     & 0.96       & 0.91     & 0.95       & 0.77     \\\hline
                            
$(1.0, \,0.1)$              & Oracle      & 0.96       & ---     & 0.96       & ---     & 0.96       & ---     & 0.96       & ---     \\
                            & Student $t$ & 1.00       & 1.69      & 0.98       & 1.20     & 1.00       & 2.19     & 0.97       & 1.25     \\
                            & IM          & 1.00       & 1.27     & 0.98       & 1.09     & 1.00       & 1.27     & 0.99       & 1.01     \\
                            & Para. Boot. & 0.98       & 1.08     & 0.97       & 1.03     & 1.00       & 1.62     & 0.96       & 1.03     \\
                            & Bayes       & 0.95       & 0.38     & 0.95       & 0.33     & 0.96       & 0.47     & 0.95       & 0.34    \\\hline
\end{tabular}%
}
\end{table}

}
\begin{table}
\centering
\caption{Observed coverage proportion and ratios of average prediction interval lengths compared to the Oracle method of $95\%$ prediction intervals for $\theta$ using the Satterthwaite approximations described in Appendix~\ref{A:satt}.}
\label{table:theta_satt}
\resizebox{1\textwidth}{!}{%
\begin{tabular}{@{}cccccccccc@{}}\hline
                            &             & \multicolumn{8}{c}{Simulation Setting}                                                        \\\cmidrule(l){3-10} 
                            &             & \multicolumn{2}{c}{A} & \multicolumn{2}{c}{B} & \multicolumn{2}{c}{C} & \multicolumn{2}{c}{D} \\
$(\sigma_a^2, \, \sigma^2)$ & Method      & Coverage   & Length   & Coverage   & Length   & Coverage   & Length   & Coverage   & Length   \\ \hline
$(0.01, \,1.0)$             & Oracle        & 0.94       & ---     & 0.96               & ---     & 0.94       & ---       & 0.96                & ---    \\
                            & Satt assoc. & 0.94       & 1.40     & \hlgray{0.92}       & 1.17     & 0.94       & 1.20     & \hlgray{0.91}       & 1.17     \\
                            & Gen Satt.   & 0.93       & 1.36     & \hlgray{0.92}       & 1.17     & 0.94       & 1.16     & \hlgray{0.91}       & 1.16     \\\hline
                            
$(0.1, \,1.0)$              & Oracle      & 0.94               & ---       & 0.95                & ---      & 0.94                & ---      & 0.95                & ---    \\
                            & Satt assoc. & \hlgray{0.85}       & 1.07     & \hlgray{0.85}       & 0.95     & \hlgray{0.84}       & 1.04     & \hlgray{0.85}       & 0.95     \\
                            & Gen Satt.   & \hlgray{0.84}       & 1.03     & \hlgray{0.85}       & 0.94     & \hlgray{0.84}       & 1.00     & \hlgray{0.85}       & 0.94     \\\hline
                            
$(0.5, \,0.5)$              & Oracle      & 0.94               & ---      & 0.95                & ---      & 0.94                & ---      & 0.94                & ---     \\
                            & Satt assoc. & \hlgray{0.88}      & 1.13     & 0.93                & 1.04     & \hlgray{0.88}       & 1.12     & \hlgray{0.92}       & 1.05     \\
                            & Gen Satt.   & \hlgray{0.86}      & 1.01     & \hlgray{0.91}       & 1.00     & \hlgray{0.85}       & 1.00     & \hlgray{0.91}       & 1.00     \\\hline
                            
$(1.0, \,0.1)$             & Oracle        & 0.95               & ---     & 0.94       & ---      & 0.94                & ---      & 0.94       & ---     \\
                            & Satt assoc.  & 0.97               & 1.34    & 0.95       & 1.10     & 0.98                & 1.40     & 0.95       & 1.13     \\
                            & Gen Satt.    & \hlgray{0.90}      & 1.09    & 0.93       & 1.06     & \hlgray{0.89}       & 1.08     & 0.93       & 1.06     \\\hline
\end{tabular}%
}
\end{table}

\begin{table}
\centering
\caption{Observed coverage proportion and ratios of average prediction interval lengths compared to the Oracle method of $95\%$ prediction intervals for a new observation $Y^\star$ in a new group using the Satterthwaite approximations described in Appendix~\ref{A:satt}.}
\label{table:new_satt}
\resizebox{1\textwidth}{!}{%
\begin{tabular}{@{}cccccccccc@{}}\hline
                            &             & \multicolumn{8}{c}{Simulation Setting}                                                        \\\cmidrule(l){3-10} 
                            &             & \multicolumn{2}{c}{A} & \multicolumn{2}{c}{B} & \multicolumn{2}{c}{C} & \multicolumn{2}{c}{D} \\
$(\sigma_a^2, \, \sigma^2)$ & Method      & Coverage   & Length   & Coverage   & Length   & Coverage   & Length   & Coverage   & Length   \\ \hline
$(0.01, \,1.0)$             & Oracle        & 0.96       & ---     & 0.96       & ---     & 0.96       & ---     & 0.96       & ---     \\
                            & Satt assoc. & 0.96       & 1.06     & 0.96       & 1.01     & 0.96       & 1.06     & 0.96       & 1.02     \\
                            & Gen Satt.   & 0.94       & 1.04     & 0.96       & 1.01     & 0.95       & 1.04     & 0.96       & 1.01     \\\hline
                            
$(0.1, \,1.0)$              & Oracle        & 0.95       & ---     & 0.96       & ---     & 0.96       & ---     & 0.96       & ---     \\
                            & Satt assoc. & 0.96       & 1.07     & 0.96       & 1.02     & 0.96       & 1.07     & 0.96       & 1.02     \\
                            & Gen Satt.   & 0.96       & 1.04     & 0.96       & 1.01     & 0.95       & 1.04     & 0.96       & 1.01     \\\hline
                            
$(0.5, \,0.5)$              & Oracle        & 0.95       & ---     & 0.95       & ---     & 0.95       & ---     & 0.95       & ---     \\
                            & Satt assoc. & 0.96       & 1.18     & 0.95       & 1.06     & 0.96       & 1.17     & 0.95       & 1.08     \\
                            & Gen Satt.   & 0.93       & 1.06     & 0.94       & 1.03     & 0.94       & 1.05     & 0.95       & 1.03     \\\hline
                            
$(1.0, \,0.1)$              & Oracle        & 0.94              & ---     & 0.94       & ---     & 0.94                 & ---     & 0.94       & ---     \\
                            & Satt assoc. & 0.98                & 1.37    & 0.95       & 1.11     & 0.99                & 1.43     & 0.96       & 1.13     \\
                            & Gen Satt.   & \hlgray{0.91}       & 1.10    & 0.94       & 1.06     & \hlgray{0.90}       & 1.10     & 0.93       & 1.06     \\\hline
\end{tabular}%
}
\end{table}

\end{document}